\documentclass{article}
\usepackage[letterpaper,top=2cm,bottom=2cm,left=3cm,right=3cm,marginparwidth=1.75cm]{geometry}
\usepackage{amsmath}
\usepackage{graphicx}
\graphicspath{ {./images/} }
\usepackage[colorlinks=true, allcolors=blue]{hyperref}
\usepackage{amsfonts}
\usepackage{physics}
\usepackage{subcaption}
\usepackage{cite}
\usepackage{mathtools}
\usepackage{setspace}
\usepackage{authblk}
\usepackage{nameref}
\title{Parametric Hypersensitivity and Transport in the Steady-State Open-System Holstein Model}
\author[1]{Nishaant Jacobus}
\author[1]{Paul Brumer}
\author[2]{Chern Chuang}
\affil[1]{Chemical Physics Theory Group, Department of Chemistry, and Center for Quantum Information and Quantum Control,
University of Toronto, Toronto, Ontario M5S 3H6, Canada}
\affil[2]{Department of Chemistry and Biochemistry, University of Nevada, Las Vegas, Las Vegas, NV 89154, USA
}
\begin{document}
\maketitle{}
\section*{Abstract}
We demonstrate that the nonequilibrium steady state (NESS) of an open-system Holstein model with linear bias displays extreme sensitivity to the closed system parameters. This sensitivity is shown to correspond to avoided crossings in the closed system spectrum, as previously demonstrated in the Rabi model. We then develop a kinetic model to analyze the effects of environmental parameters on NESS hypersensitivity. This reveals that hypersensitivity only exists in intermediate environmental parameter regimes, a prediction that is verified numerically. The inherent spatial character of the Holstein model offers a natural connection to transport, revealing that transport properties in the steady-state regime can be optimized by simultaneously coordinating the closed- and open-system parameters.
\section*{I. Introduction}
The behavior of open quantum systems is of great interest in many areas such as chemical dynamics, biological systems, and quantum computing \cite{nitzan_chemical_2014, mohseni_quantum_2014, benenti_principles_2018}. Due to environmental effects such as dissipation and driving, these systems can display nonequilibrium steady states (NESS) that differ significantly from the closed system dynamics \cite{hershfield_reformulation_1993, anders_steady-state_2008, hsiang_nonequilibrium_2015, lotem_renormalized_2020}. Furthermore, our previous computational studies of the photoisomerization of model rhodopsin (the first step of vision) \cite{chuang_extreme_2021, chuang_steady_2022} and the Rabi model, a two-level system coupled to a bosonic mode \cite{chuang_parametric_2023}, show that these NESS can display extreme sensitivity to system parameters, a behavior not seen in the transient dynamics. Our study of the Rabi model also revealed that this sensitivity is connected to avoided crossings in the closed system spectrum \cite{chuang_parametric_2023}.
\newline
\newline Here we extend the analysis of parametric hypersensitivity, showing this behavior in the Holstein model, a model for electron-phonon or exciton-phonon coupling that has been applied to systems such as polymers and semiconductors \cite{kessing_long-range_2022, coropceanu_charge_2007}. In the Holstein model each site (exciton) is coupled to a local bosonic mode, providing additional degrees of freedom not seen in the Rabi model; here we show that the resulting system displays new kinds of NESS sensitivity patterns. As was the case for the Rabi model, these open-system resonances are shown to be understood in terms of avoided crossings in the underlying closed system; they also correspond to the enhanced closed-system transport properties that were discussed in Ref \cite{kessing_long-range_2022}, but now assisted by dissipative effects. This paper therefore extends the findings of \cite{kessing_long-range_2022}, which focused on closed-system dynamics, to the open system in the steady state. 
\newline
\newline The one-dimensional nature of the Holstein model provides an inherent spatial measure that quantifies the extent of resonance enhancement in transport, absent in our previous Rabi study. This leads us to develop, as discussed in this paper, a new understanding of resonance hypersensitivity in terms of the kinetics of population transfer. These results constitute a major step towards understanding the range of generality of parametric sensitivity in molecular systems in the NESS.
\newline
\newline The kinetic approach introduced below also allows us to understand the effects of varying environmental parameters on the NESS populations. We find, significantly, that hypersensitivity can only manifest under certain environmental conditions, and that this is linked to the effect of the environment on exciton transport kinetics in the condensed phase. Specifically, the environment is found to affect both the rate of nearest-neighbor transport and resonance-enhanced long-range ``jump" transport; in environmental regimes where both of these processes are either too fast or too slow the resonance-dependent NESS population distribution ceases to exist. Hypersensitivity only appears in intermediate regimes where the nearest-neighbor transport is slow and where the resonance jump transport provides a sufficiently fast alternative pathway.
\section*{II. Model}
\begin{figure}[h]
\includegraphics[width=\linewidth]{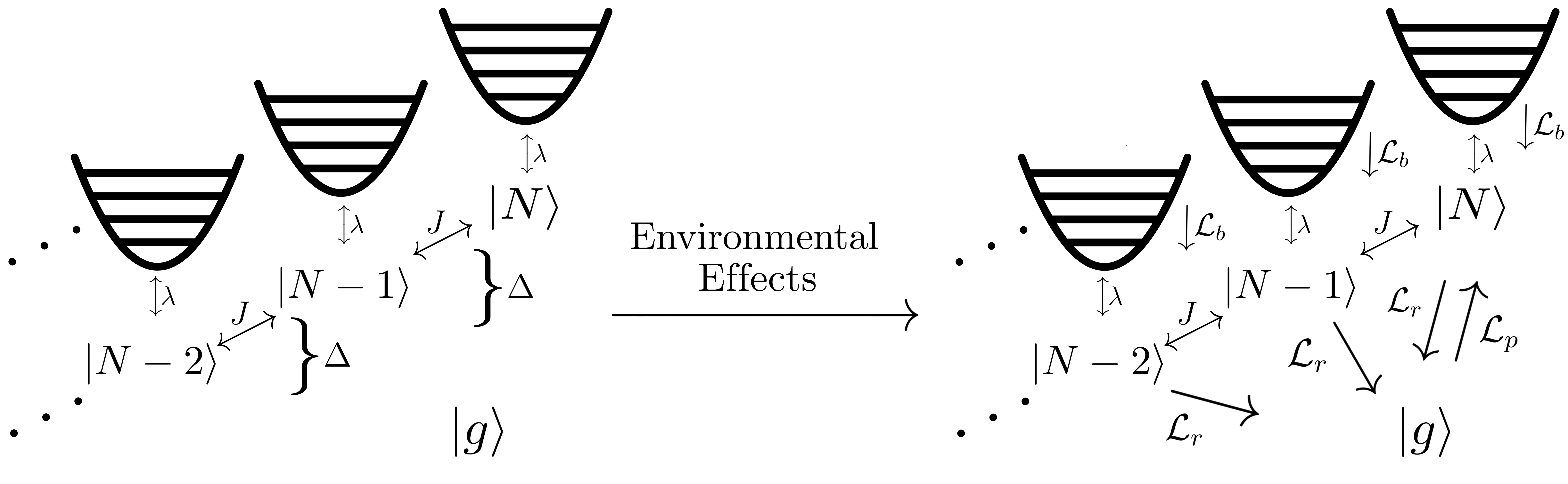}
\centering
\caption{The biased Holstein chain model described by the Hamiltonian (\ref{closed_sys_hamiltonian}), before and after the addition of open system dynamics described by equations (\ref{lindblads}).}
\centering
\label{open_and_closed_schematic}
\end{figure}

For an $N$-site chain where each site is coupled to a local bosonic mode, the Holstein model Hamiltonian can be written as:
\begin{equation}
\begin{split}
    H_\textrm{sys} &= H_\textrm{site} + H_\textrm{boson} + H_\textrm{site-boson} \\
    &= \left[\sum_{n = 1}^N \varepsilon_n c^\dagger_nc_n+ \sum_{\langle n, m \rangle}\left( J_{n,m} c_n^\dagger c_m + \textrm{H.c.}\right)\right] + \sum_{n = 1}^N \omega_n a_n^\dagger a_n + \sum_{n = 1}^N \lambda_n c_n^\dagger c_n (a_n^\dagger + a_n),
\end{split}
\end{equation}
where $c_n^\dagger \ (c_n)$ denotes an exciton creation (annihilation) operator for the $n$th site, $a_n^\dagger \ (a_n)$ denotes a bosonic creation (annihilation) operator for the $n$th mode, $J_{n,m}$ denotes the site-to-site coupling, H.c. denotes Hermitian conjugate, and $\varepsilon_n, J_{n,m}, \omega_n, \lambda_n \in \mathbb{R}$. With a linear bias and identical sites with only nearest-neighbor coupling, the system Hamiltonian can be written (in units of bosonic energy $\omega$) as:
\begin{equation} \label{closed_sys}
    H_\textrm{sys} = \varepsilon_{10} + \sum_{n = 1}^N \left\{\left[n\Delta + \lambda \left(a_n^\dagger +  a_n\right)\right] c_n^\dagger c_n + a_n^\dagger a_n\right\} + J \sum_{n=2}^N \left(c_n^\dagger c_{n-1} + \textrm{H.c.}\right)
\end{equation}
where $\varepsilon_{10}$ is an overall energy gap. The chain is labeled so that $\Delta \ge 0$ (i.e. $n = 1$ is the lowest-energy site and $n = N$ is the highest).
\newline
\newline We restrict attention to the ground (zero-) and one-exciton subspace of the total Hilbert space, denoting the ground state as $|g \rangle$, and using the product basis for the one-exciton space. The state that has the exciton located on the $n$th site and the $j$th bosonic mode in its $m_j$th excited state is denoted $|n; m_1, m_2, ..., m_N \rangle$, (note that in this convention the ground state of the bosonic mode is $m_j = 0$). The energy of $|g \rangle$ is set to 0 and the one-exciton Hamiltonian is written as:
\begin{equation} \label{closed_sys_hamiltonian}
 H_\textrm{sys}^{(1)} = \varepsilon_{10} +  \sum_{n = 1}^N (n\Delta + \lambda (a_n^\dagger +  a_n)) |n \rangle \langle n| +  a_n^\dagger a_n +  \sum_{n = 2}^N (J |n \rangle \langle n - 1 | + \textrm{H.c.}).
\end{equation}
Note that although the closed system Hamiltonian does not couple the ground state to the manifold of one-exciton states, the ground state is included in order to describe dissipation and excitation dynamics in the open system, discussed later below.
\newline
\newline The process of interest comprises constant incoherent pumping into the highest-energy site, followed by redistribution of energy and populations amongst lower energy sites. This is modeled using a Lindbladian formalism, where the time evolution for the density operator $\rho$ is given by:
\begin{equation} \label{open_sys_eom}
\begin{split} \dot{\rho}(t) &= -i[H_\textrm{sys}, \rho(t)] + \left(\sum_{n = 1}^N \mathcal{L}_{b,n} + \sum_{n = 1}^N \mathcal{L}_{r, n} + \mathcal{L}_p\right)\rho(t),
\end{split}
\end{equation}
and where the Liouvillian superoperators are defined as:
\begin{equation} \label{lindblads}
\begin{split}
\mathcal{L}_y \rho(t) = L_y \rho(t) L_y^\dagger - \frac{1}{2} \{L_y^\dagger L_y, \rho(t) \},
\end{split}
\end{equation}
with each $L_y$ representing a system-bath interaction. For each $n \in \{1, ..., N\}$, $L_{r,n} = \sqrt{r_r}c_n$, where $r_r$ is the inverse lifetime of the exciton. $L_{b,n} = \sqrt{r_b} a_n$, so that $L_{b,n}$ represents the bosonic dissipation of the $n$th mode with inverse lifetime $r_b$. Finally $L_p = \sqrt{r_p} c_N^\dagger$, representing an incoherent pumping process into the highest-energy site $N$. The pumping process is of the slowest time scale to justify a restriction to the zero- and one-exciton manifolds, which also implies that the bosonic dynamics in the ground state is irrelevant. This model is illustrated in Fig. \ref{open_and_closed_schematic}.
\section*{III. Closed System Spectrum and NESS Hypersensitivity}
\subsection*{Avoided Crossings and Eigenstate Spatial Distributions}
\begin{figure}[h]
\includegraphics[width=0.75\linewidth]{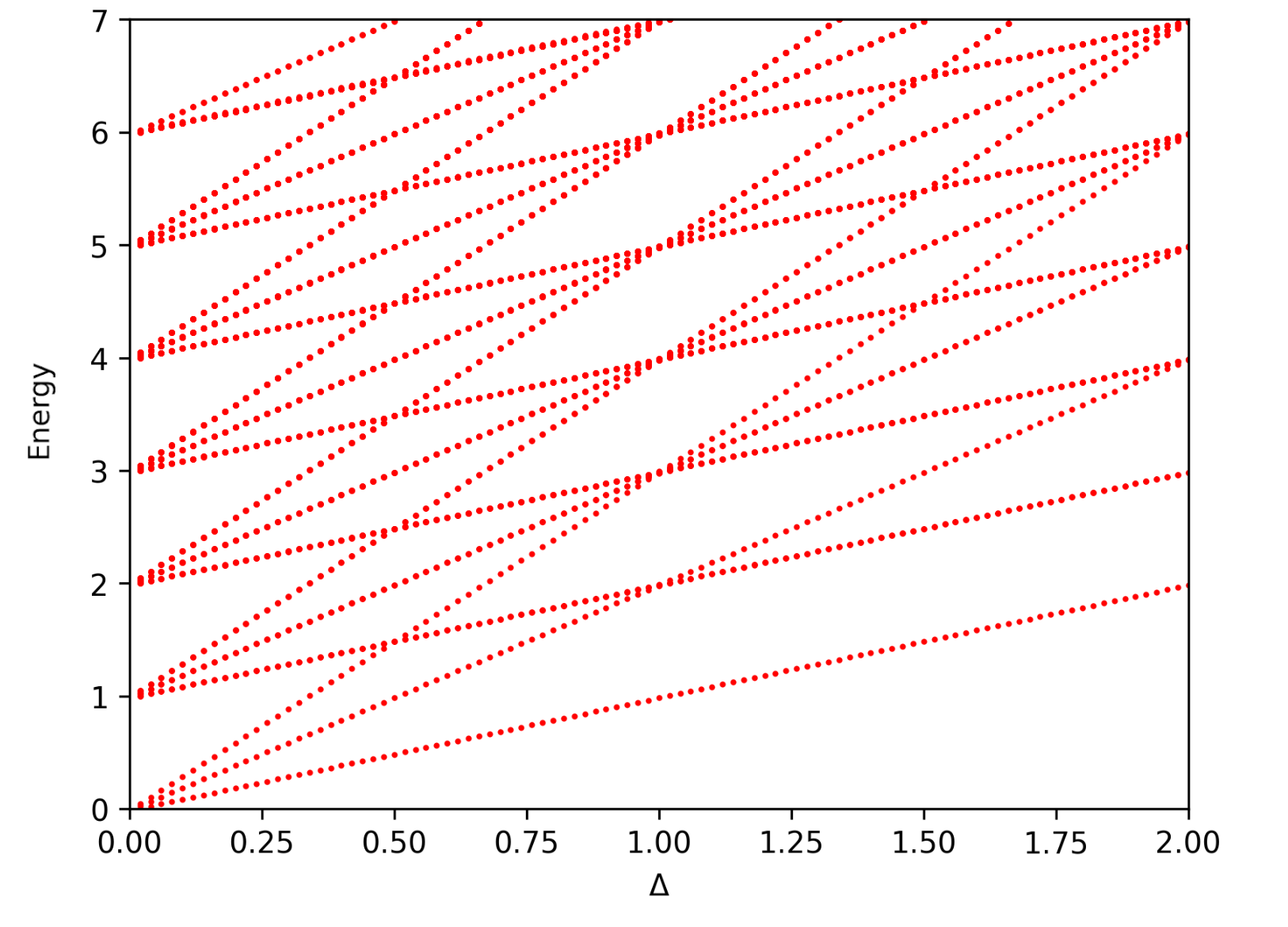}
\centering
\caption{Numerically calculated lowest one-exciton state energies of a Holstein trimer for $\lambda = 0.4$ and $J = 0.01$. A three-fold branching occurs due to the $k\Delta$ term in (\ref{closed_sys_energy}), and each set of branches corresponds to a different total bosonic excitation. Avoided crossings occur at integer and half-integer $\Delta$.}
\centering
\label{holstein_trimer_spectrum}
\end{figure}

To investigate whether the connection between avoided crossings and NESS hypersensitivity found in \cite{chuang_parametric_2023} for the Rabi model also holds for the Holstein Model, it is first necessary to resolve the closed system spectrum. We restrict to the strong bias regime, $J/\Delta \ll 1$. To isolate the effects of exciton-phonon coupling, we transform from the site basis states $|j\rangle_{s}$ to the Wannier-Stark (WS) basis $|j\rangle_{WS}$ defined by \cite{zhang_holstein_2002}:
\begin{equation}
\label{ws_def}
    |j\rangle_{WS} = \sum_{i = -\infty}^\infty J_{i-j}\left(\frac{2J}{\Delta}\right)|i\rangle_s
\end{equation}
where $J_n(z)$ is the Bessel function of the first kind (note $J$ with a subscript denotes a Bessel function while $J$ without a subscript is the nearest neighbor coupling strength). Notice the strong bias localizes the $j$th WS state around site $j$ (Wannier-Stark localization \cite{wannier_dynamics_1962, bleuse_electric-field-induced_1988, mendez_stark_1988}). Letting $d_j^\dagger$ denote the creation operator for the $j$th WS state, the  Hamiltonian (\ref{closed_sys}) becomes \cite{zhang_holstein_2002}:
\begin{equation}
\label{H_ws}
    H_{sys} = H_{tb} + V
\end{equation}
with the diagonalized tight binding Hamiltonian given by:
\begin{equation}
H_{tb} = \varepsilon_{10} + \sum_{j = -\infty}^\infty j\Delta \left(d_j^\dagger d_j\right) + a_j^\dagger a_j
\end{equation}
and the coupling $V$ given by:
\begin{equation} \label{WS_state_coupling}
V = \lambda \sum_{i,j,k} J_{j - i}\left(\frac{2J}{\Delta}\right)J_{j - k}\left(\frac{2J}{\Delta}\right)(a_j^\dagger + a_j)d^\dagger_id_k
\end{equation}
This form of the Hamiltonian makes the long-range couplings between the WS states (and hence approximately between the sites) manifest. Specifically, the long-range coupling is mediated by the exciton-phonon coupling term and attenuated by the exponentially decaying overlap between WS wavefunctions, as will be seen below. The strong localization of the WS states also suggests that the WS basis should approximately diagonalize the tight-binding terms of (\ref{closed_sys}) when the chain length $N$ is finite, so that we can use the form (\ref{H_ws}) to analyze finite chains as well.
%, so these considerations about infinite chains should apply to finite ones as well. 
\newline
\newline Starting with the unperturbed Hamiltonian $H_{tb}$, the direct product state $|k\rangle_{WS}\otimes|\vec{m} \rangle = | k; m_1, ..., m_N \rangle_{WS}$ has energy:
\begin{equation} \label{closed_sys_energy} E_{k,\vec{m}} = k \Delta + \sum_{j=1}^N m_j = k\Delta + n_b
\end{equation}
States that are both in the same WS state and have the same total number of bosonic excitations $n_b= \sum_{j=1}^N m_j$ are degenerate, so in general eigenstates can be labeled by the WS site number $k$ and by the number of bosonic quanta. Eq. ($\ref{closed_sys_energy}$) indicates that states located on WS sites $k$ and $k + m$ can only be degenerate when $\Delta = j/m$ for some integer $j$ (with $\Delta = j/m$ in reduced form). Physically, this condition corresponds to an extra $j$ bosons ``filling in" the energy gap between the WS states, analogous to a vibrational manifold that fills in the energy gap between two electronic states. Henceforth, a $\Delta$ value is termed resonant if it has such a degeneracy between different WS states.
\newline
\newline Upon perturbing $H_{tb}$ with the exciton-phonon coupling $V$, one would anticipate the formation of avoided crossings at the resonant $\Delta$ due to the presence of coupling between different WS states in $V$. Noting that the strength of the coupling between WS states $i$ and $k$ via phonon mode $j$ is characterized by $|J_{j - i}\left(2J/\Delta\right)J_{j-k}\left(2J/\Delta\right)|$, and using the asymptotic representation of Bessel functions near the origin \cite{abramowitz_handbook_1964}:
\begin{equation} \label{bessel_asymptote}
J_n(z) \approx \begin{cases}
			\frac{1}{n!}\left(\frac{z}{2}\right)^n, & \text{$n\ge0$}\\
            \frac{(-1)^n}{(-n)!}\left(\frac{z}{2}\right)^{-n}, & \text{$n<0$}
		 \end{cases}
\end{equation} we see that $|J_{j - i}\left(2J/\Delta\right)J_{j - k}\left(2J/\Delta\right)|$ scales as $(J/\Delta)^{|j-i| + |j - k|}$. By ranging over all phonon modes $j$, the hopping strength between sites $i$ and $k$ is seen to scale as $\lambda(J/\Delta)^{|i-k|}$ to leading order; the coupling consequently drops exponentially with the distance between sites. These considerations are validated numerically in Fig. \ref{holstein_trimer_spectrum}, which shows the numerically calculated spectrum of a Holstein trimer ($N = 3$). For numerical implementation, each bosonic mode is truncated at the 10th level, which was found sufficient for convergence for the parameter regimes considered; the QuTiP library was also used for some of the calculations \cite{johansson_qutip_2013}. At $\Delta = 0$ there are sets of levels spaced by one bosonic energy quantum, which correspond to different total bosonic excitations. As $\Delta$ increases, these levels split into three prongs with integer slopes, with one prong correspond to each occupied site (an $N$-mer would have $N$ prongs), reflecting the unperturbed energies given by Eq. (\ref{closed_sys_energy}). Avoided crossings occur at integer and half-integer $\Delta$, although not visually evident in the figure. The exponential dropoff in coupling strength leads to a narrower splitting in half-integer avoided crossings than integer ones.
\subsection*{Hypersensitivity}
\begin{figure}[h]
\includegraphics[width=0.75\linewidth]{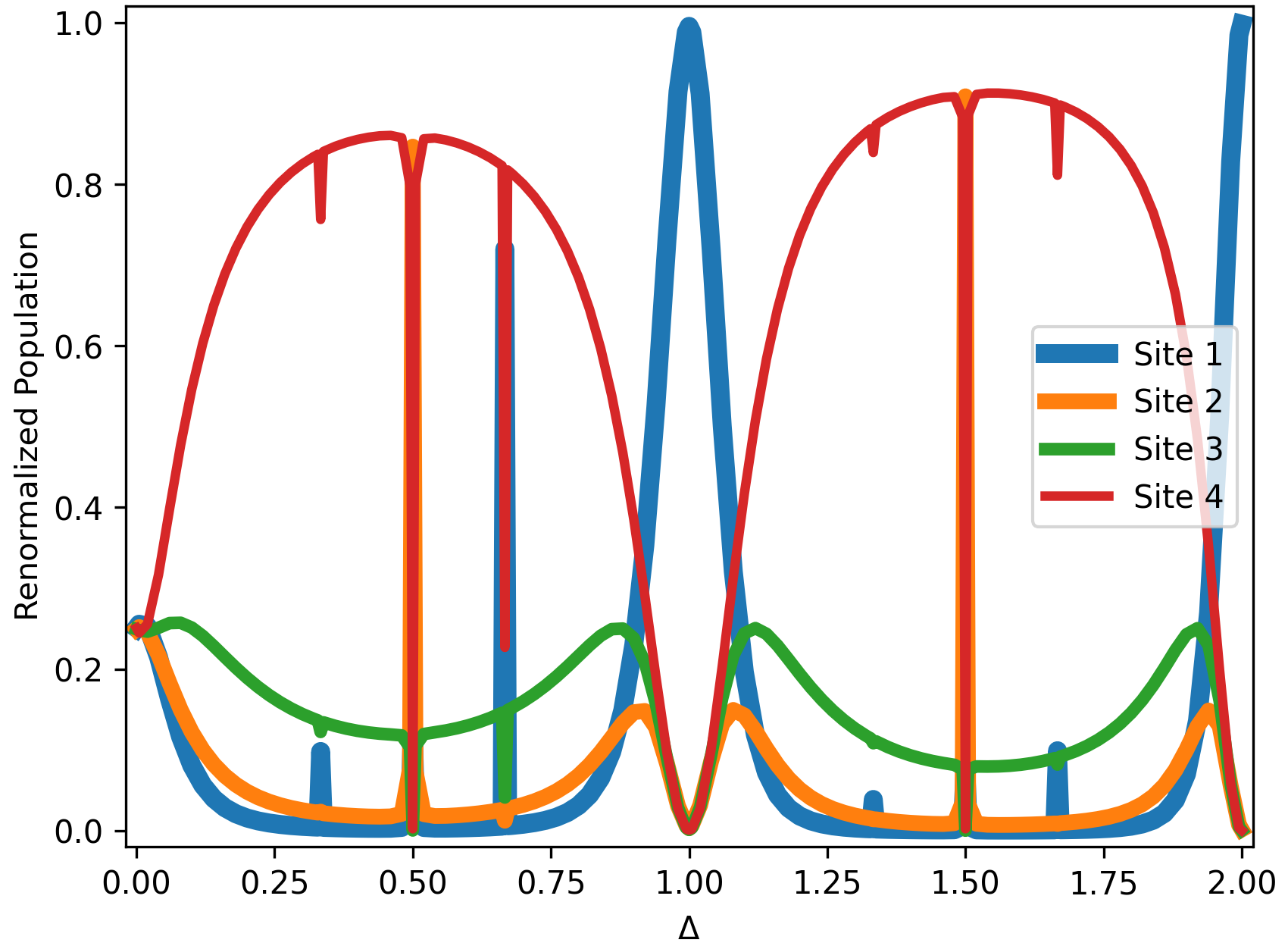}
\centering
\caption{Numerically calculated normalized one-exciton site populations for a tetramer model; $\lambda = 0.5, J = 0.01, r_b = 10^{-7}, r_r = 10^{-9}, r_p = 10^{-10}$. Populations are normalized by dividing by the total one-exciton population and sum to unity. Site two has a spike in population at half-integer resonances, while site one has a spike in population at the $\Delta = 2/3$ fractional resonance, showing that only the sites connected by the jump mechanism are affected by the resonance.}
\centering
\label{holstein_tetramer_NESS}
\end{figure}

\begin{figure*}[h!]
    \centering
    \begin{subfigure}[t]{0.5\textwidth}
        \centering
        \includegraphics[width=7.6cm]{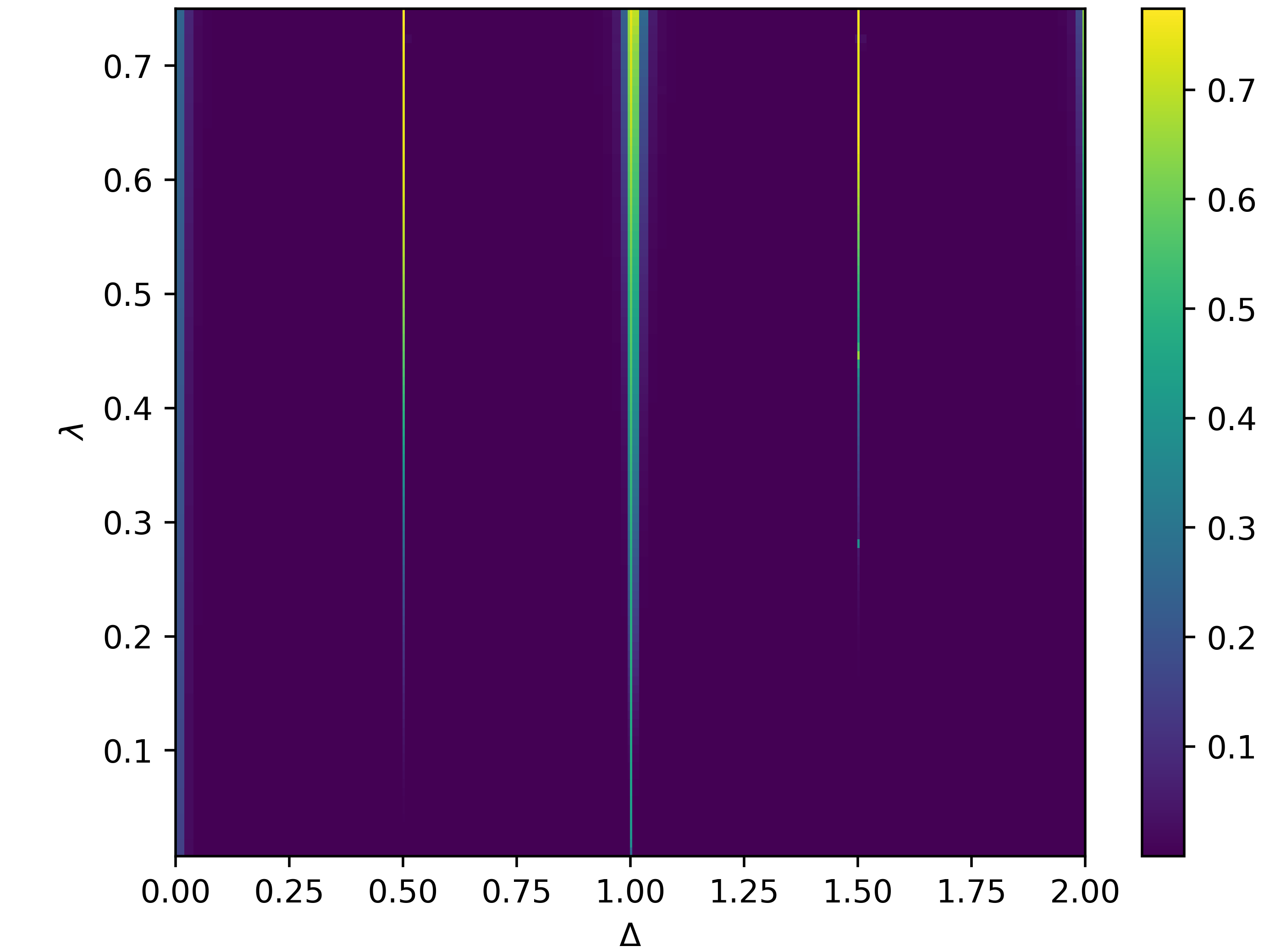}
        \caption{}
        \label{lambda_indep_spike_loc}
    \end{subfigure}%
    ~ 
    \begin{subfigure}[t]{0.5\textwidth}
        \centering
        \includegraphics[width=7.5cm]{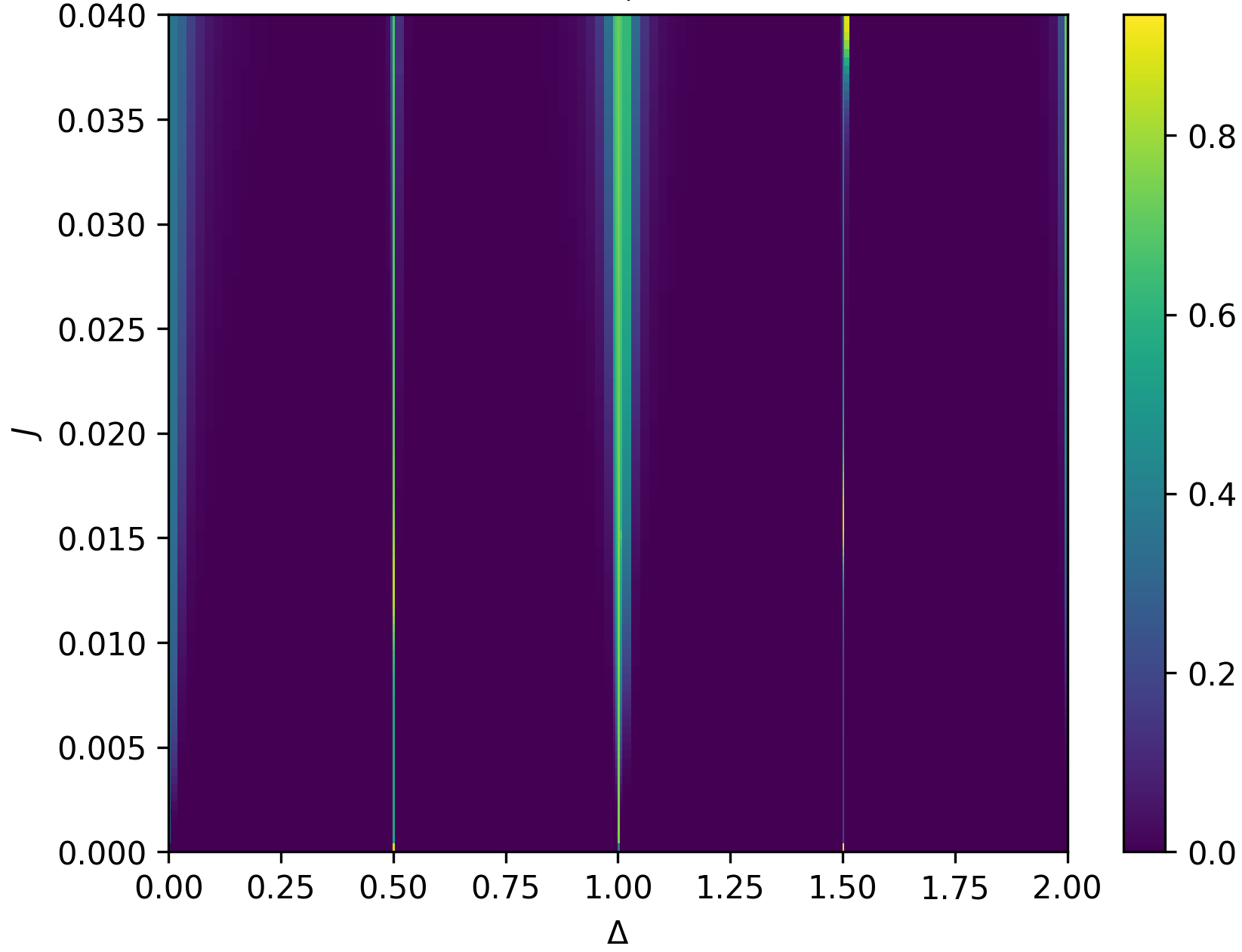}
        \caption{}
        \label{J_indep_spike_loc}
    \end{subfigure}
    \caption{Normalized bottom site population for a Holstein trimer as a function of $\Delta$ (horizontal axis) and (a) $\lambda$, with $J$ held fixed at $0.01$, and (b) $J$, with $\lambda$ held fixed at $0.4$. In all calculations, $r_b = r_r = 10^{-9}, r_p = 10^{-10}$. The population spikes appear at half-integer and integer resonances regardless of the $\lambda$ and $J$ values, but become wider with increasing $\lambda$ and $J$, reflecting that the coupling strength scales as $\lambda(J/\Delta)^m$ for $m$-length jumps.}
    \label{lam_and_J_NESS_independence}
\end{figure*}

With the avoided crossings determined to occur at appropriate integer and fractional $\Delta$, we now look for the corresponding NESS hypersensitivity. The steady state is obtained numerically by solving for the kernel of (\ref{open_sys_eom}) within the secular approximation (the validity of which is discussed in Appendix \ref{appen_secular}), in which the populations $P$ of the energy eigenbasis evolve according to:
\begin{equation}
    \frac{\partial P^{(e)}}{\partial t} = \sum_y W_y P^{(e)}
\end{equation}
with:
\begin{equation} \label{secular_EOM}
    \langle m^{(e)}| W_y | n^{(e)} \rangle = |L_{y, mn}^{(e)}|^2 - \delta_{mn}\sum_{k} |L_{y, kn}^{(e)}|^2.
\end{equation}
Here $(e)$ denotes the energy eigenbasis, and the energy eigenbasis coherences are assumed negligible in the NESS (note, however, \cite{dodin_population_2024}). In anticipation of a later section it is useful to present results in terms of normalized one-exciton populations, $\rho_n \coloneqq P_n/P_{ex}$, where $P_n$ is the population on the $n$th site and $P_{ex} = \sum_{n = 1}^N P_n$ is the total excited state population. Numerical calculations indeed reveal hypersensitivity associated with the avoided crossings. As an example, the $\Delta$-dependence of NESS normalized site populations of a tetramer model ($N = 4$) is shown in Fig. \ref{holstein_tetramer_NESS}. For most off-resonance $\Delta$ values, the normalized NESS population is almost entirely concentrated on the top site (red). However, when $\Delta = 1/2$ or $3/2$, there is a spike in population on site 2, corresponding to an avoided crossing between states localized on site 2 and the top site. Similarly, when $\Delta = j/3$ for $j = 1,2, 4, 5$ there is increased population of site 1, matching the avoided crossing between states localized on sites 1 and the top site. In general, for avoided crossings at non-integer fractional $\Delta$, the sites are selectively populated in the NESS according to which states have avoided crossings with top site states. This contrasts with the integer $\Delta$ case, where avoided crossings involve states localized on every site, and in which significant population is often seen on all of the sites. A dropoff in peak prominence is also seen with jump length, with the $\Delta = j/3$ resonances being much weaker. This also reflects the dropoff in the coupling with increasing site distance. In particular, this exponential dropoff means that in the large $N$ limit, we will not see significant hypersensitivity at all rational $\Delta$, even though all rational $\Delta$ become resonant values in principle.
\newline
\newline As a further indication that this NESS behavior corresponds to avoided crossings, Fig. \ref{lam_and_J_NESS_independence} shows the numerically calculated population of the bottom site for a trimer when $\lambda$ and $J$ are varied; the resonances seem to be independent of $\lambda$ or $J$ (within the perturbative $J$ regime). This is expected, as Eq. (\ref{closed_sys_energy}) indicates that the occurrence of avoided crossings depends on $\Delta$ alone. However, we do see the peaks broaden with increasing $J$ and $\lambda$, reflecting that the spectral gap has to be compared to $J$ and $\lambda$ since these are what the perturbative exciton-phonon coupling depends on.
\newline
\newline %We briefly note that our choice to look for resonances in the normalized populations instead of the actual excited populations is justified because the transitions between the ground and excited state are controlled only by the environmental parameters $r_p$ and $r_r$ and make no reference to the spatial location of the exciton. This suggests that the total excited population $P_{ex}$ should remain approximately constant as $\Delta$ is varied, meaning that the normalized and total populations just vary by an overall constant and therefore show the same resonances. We find this holds numerically as well, with the plot of the unnormalized excited populations as a function of $\Delta$ having the same resonance features as the plot of normalized populations. However, in section \hyperref[bath_parameter_section]{V.} we will vary the environmental parameters in order to determine environmental effects on sensitivity; to make fair comparisons between different regimes it will be necessary to use the normalized populations. 
\section*{IV. Kinetic Perspective on Hypersensitivity}
\subsection*{Secular Kinetics and Spatial Distribution of Eigenstates}
\begin{figure*}[h!]
    \centering
    \begin{subfigure}[t]{0.5\textwidth}
        \centering
        \includegraphics[width=7.5cm]{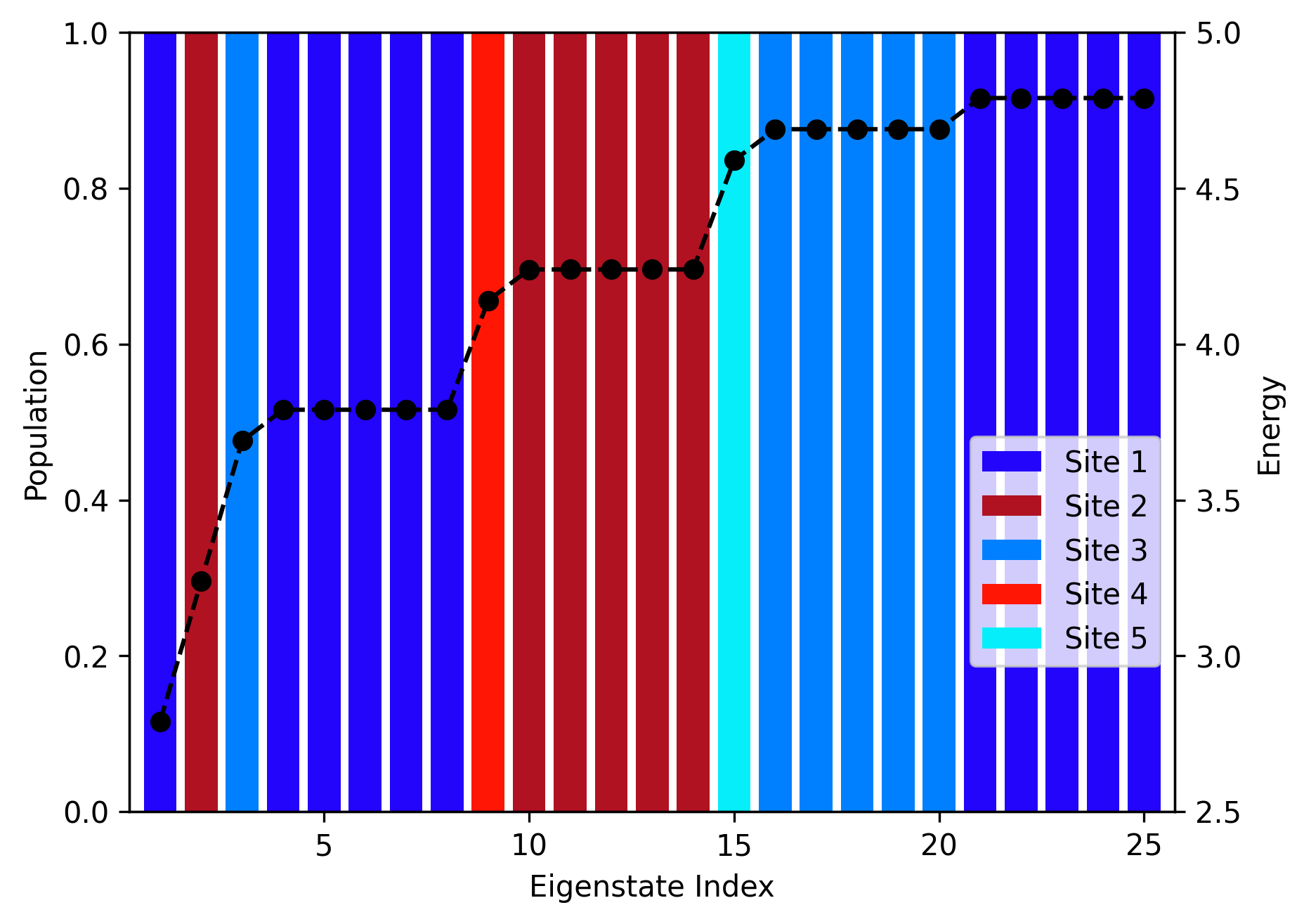}
        \caption{}
        \label{off_res_localization}
    \end{subfigure}%
    ~ 
    \begin{subfigure}[t]{0.5\textwidth}
        \centering
        \includegraphics[width=7.5cm]{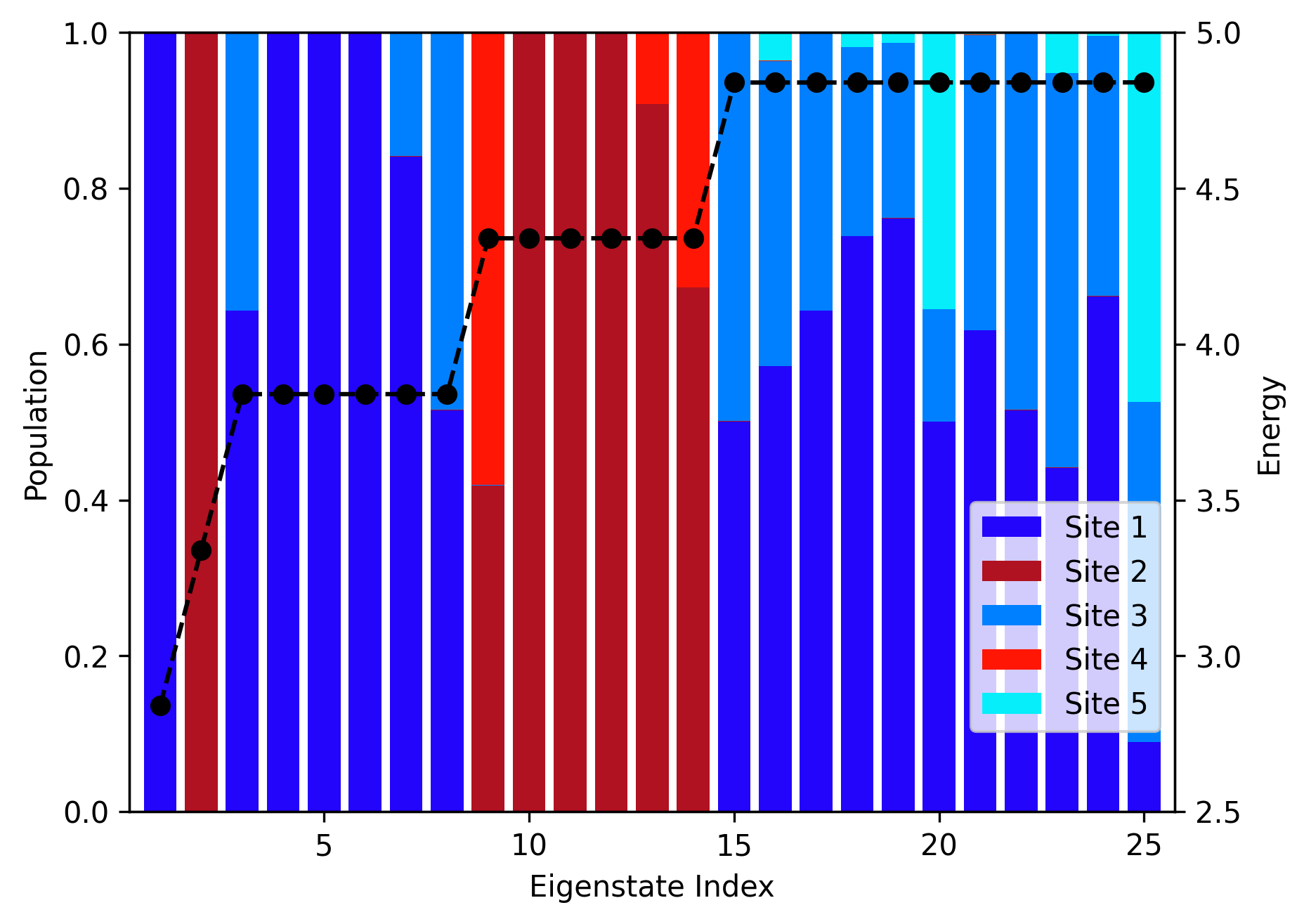}
        \caption{}
        \label{on_res_delocalization}
    \end{subfigure}
    \caption{Numerically calculated site populations of the lowest one-exciton eigenstates for a five-site system with (a) $\Delta = 0.45$ (off-resonance) and (b) $\Delta = 0.50$ (on-resonance). All calculations are done for $\lambda = 0.4$ and $J = 0.01$. The superimposed dashed line shows the eigenstate energies; degeneracy only occurs between states on the same site in the off-resonance case, but states on alternating sites become degenerate at the $\Delta = 0.5$ resonance, leading to delocalized eigenstates. To better illustrate the mixing pattern, the odd sites are blue and the even sites are red.}
    \label{site_localization}
\end{figure*}

 The previous section illustrates that the Holstein Model's NESS sensitivity can be understood based on avoided crossings; here we show that it can also be understood via a kinetic approach that incorporates the effects of the environment. This will be useful in the next section that investigates the effects of environmental parameters on hypersensitivity. The key to the kinetic approach is to recall that secular dynamics decouple the population and coherence evolution in the energy eigenbasis, and that the closed-system Hamiltonian only contributes to evolution by determining the identity of the eigenbasis, as seen by Eq. (\ref{secular_EOM}). Therefore, all closed-system effects on the dynamics and hence on the NESS (including the onset of hypersensitivity) are consequences of the spatial profiles of the eigenstates. This also hints at why hypersensitivity occurs at the avoided crossings: they are associated with rapid variation in the identity of the energy eigenstates due to the onset of strong mixing. This reasoning assumes the validity of a secular treatment, which should be sufficient to determine the dynamics relevant for the steady-state so long as any long-lasting coherences are negligible
 (the validity of the secular approximation is further discussed in Appendix \ref{appen_secular}).
\newline
\newline Consider the form of environment-induced kinetics, as governed by the Lindbladians $\mathcal{L}_{b,n}, \mathcal{L}_{r,n}$, and $\mathcal{L}_p$. Since our focus is on understanding the NESS hypersensitivity, which involves changes in the distributions of the exciton population along the chain, the operators of interest are the $\mathcal{L}_{b,n}$, because  $\mathcal{L}_p$ and the $\mathcal{L}_{r,n}$ only create/uniformly dissipate excitons and therefore play a less important role in the relative distribution of excitons along the chain. Because the $\mathcal{L}_{b,n}$ only act on the bosonic mode, they can only induce coupling between two eigenstates $|m\rangle$ and $|p\rangle$ if there is spatial overlap in the exciton coordinate. Therefore, to gain insights into the kinetics leading to NESS hypersensitivity, it is important to understand how the perturbation $V$ affects the spatial characteristics of the eigenstates. Due to the exponential damping of coupling strength, coupling between different WS states can be neglected when $\Delta$ is not within the vicinity of a resonance. Consequently, for these $\Delta$, $V$ can not affect the spatial profile of the states: the exciton remains centered on a fixed site with a WS state profile. From the definition of the WS state (\ref{ws_def}) and the asymptotic form of the Bessel function (\ref{bessel_asymptote}), we see that, to leading order in $J/\Delta$, a WS state only overlaps with other WS states localized on adjacent sites; consequently, exciton transport induced by $\mathcal{L}_{b,n}$, which scales as the square of the spatial overlap, only occurs between adjacent sites. Therefore for off-resonant $\Delta$ one finds weak bath-induced nearest-neighbor exciton transport, with excitons ``flowing" site by site due to the bath coupling.
\newline
\newline However, as $\Delta$ approaches a resonant value, the gap between states localized on different sites with appropriate amounts of bosonic excitations (determined by the boson-assisted energy-matching condition) become comparable to $J$, so that the long-range coupling between those states is no longer perturbative. This allows for strong mixing that yields eigenstates highly delocalized over specific WS states. Notably, this ``re-delocalization" \cite{zhang_holstein_2002} process only occurs over certain sites and skips others, as determined by the resonant $\Delta$ value and energy-matching condition. For example, at half-integer $\Delta$ values, excitons are delocalized across every other site, at $\Delta = j/3$ they are delocalized over every three sites, and so on. To illustrate the re-delocalization effect, the spatial probability distributions of the numerically calculated lowest energy  eigenstates for a pentamer ($N = 5$) are shown in Fig. \ref{site_localization}. When $\Delta = 0.45$ (Fig. \ref{off_res_localization}), there is a zero-boson eigenstate localized on each site (eigenstates 1-3, 9, and 15), as well as collections of states localized on a single site, such as the five-fold collection of states localized on site 1 (eigenstates 4-8); these latter states are the $N$-fold degenerate one-boson states. 
%Generally with $n_b$ bosonic quanta the degeneracy for each site-localized state is the same as the multiplicity of an $N$-mode Einstein solid with $n_b$ excitations, given by\cite{schroeder_introduction_2021}:
%\begin{equation} \label{degeneracy_expression}
%\frac{(N + n_b - 1)!}{N!(n_b - 1)!}.\end{equation}
When $\Delta = 0.50$ (Fig. \ref{on_res_delocalization}), minibands form and a systematic delocalization occurs according to the spectral degeneracies: the lowest two levels, corresponding to the zero-boson states localized on site 1 and 2, remain localized because they are energetically separated from other states, while most of the higher energy levels delocalize across sites located 2, 4, ... positions away. Similarly, eigenstates 3-8 only involve delocalization between sites 1 and 3, while a full site 1-3-5 delocalization can only occur for eigenstates 15 and higher, where we find site 1 states with at least two bosons, necessary to match the energy of a site 5 state. 
%More generally, Figure \ref{site_localization} shows that the proportions of populations on a given site is the same before and after a set of eigenstates is mixed; for example, in Figure \ref{on_res_delocalization}, the total height of the dark blue bars in eigenstates 3-8 is five, while the total height of the medium blue bars is one, representing that five site 1 states were mixed with one site 3 state. This observation holds generally, and follows from the fact that the mixing at each avoided crossing can be represented by a unitary transformation involving only the degenerate states prescribed by the energy-matching condition. Combining this with the expression (\ref{degeneracy_expression}) allows one to quantify the relative contributions from each site to the mixing that occurs at any avoided crossing. 
\newline
\newline At resonant $\Delta$, re-delocalization has a drastic effect on the kinetics: due to the large long-range mixing, $\mathcal{L}_{b,n}$ can now strongly couple sites connected by the energy-matching condition via overlaps between delocalized eigenstates. Notably, the highly structured delocalization translates into a very specific bath-induced long-range transport: it induces transport every two sites in the half-integer $\Delta$ case, every three sites when $\Delta = j/3$, and so on.  One would also expect that when this long-range coupling is present, it is much stronger than the nearest-neighbor coupling because it arises, at least for the small denominator resonances, due to strongly spatially mixed eigenstates, whereas the nearest-neighbor coupling arises due to only weak non-degenerate mixing on the order of $J/\Delta$. We note also that the long-range transport occurring here is of a different origin than the closed-system long-range transport discussed in \cite{kessing_long-range_2022}: there, the long-range coupling arises through the nearest-neighbor coupling $J$ term. By contrast, in the open system, the long-range coupling arises via the bosonic dissipation Liouvillian. However, ultimately it is the energy matching condition and corresponding resonance eigenstate delocalization that is responsible for both types of long-range couplings occurring at only fractional $\Delta$ values.
\newline
\newline In addition to the existence of the long-range transport at resonant $\Delta$, it is also important to note that, due to the dissipative nature of $\mathcal{L}_{b,n} = \sqrt{r_b}a_n$, the transport it induces (both nearest-neighbor and long-range) is highly directional. Each application of $\mathcal{L}_{b,n}$ lowers the total bosonic energy by one, forcing the exciton to populate lower-energy states. While there can be some lower-energy states higher up the chain, the lower-energy states predominantly consist of states lying in the lower-energy WS states localized towards the lower part of the chain (note this assumes $\Delta$ is not small). This implies that both the nearest-neighbor and long-range bath coupling induced by $\mathcal{L}_{b,n}$ exhibit strong directionality, promoting downwards population flow. 
%From a dynamical point of view, this resonance degeneracy would lead to the resonant closed-system long-range transport discussed in \cite{kessing_long-range_2022}, with the long-range perturbative couplings only inducing significant transport at degeneracies.% because the degeneracy means there is no oscillatory $e^{i\omega t}$ term \cite{cohen-tannoudji_quantum_1986} in the interaction picture coupling matrix element between the states, allowing for a large long-range transfer of population. 
%Because of the exponential decrease in coupling strength with distance, the mixing between degenerate states at fractional resonant $\Delta$ becomes weaker when the denominator of $\Delta = j/m$ increases (since $\Delta = j/m$ induces mixing between sites $m$ away), making the resonance eigenstates more localized. Equivalently, in the closed-system time-dependent picture this weaker coupling is reflected in the decreased prominence of resonant transport peaks at higher denominator $\Delta$, which is what is seen in \cite{kessing_long-range_2022}. 
\subsection*{Kinetic Model} 
\begin{figure*}[h!]
    \centering
    \begin{subfigure}[t]{0.5\textwidth}
        \centering
        \includegraphics[width=7.5cm]{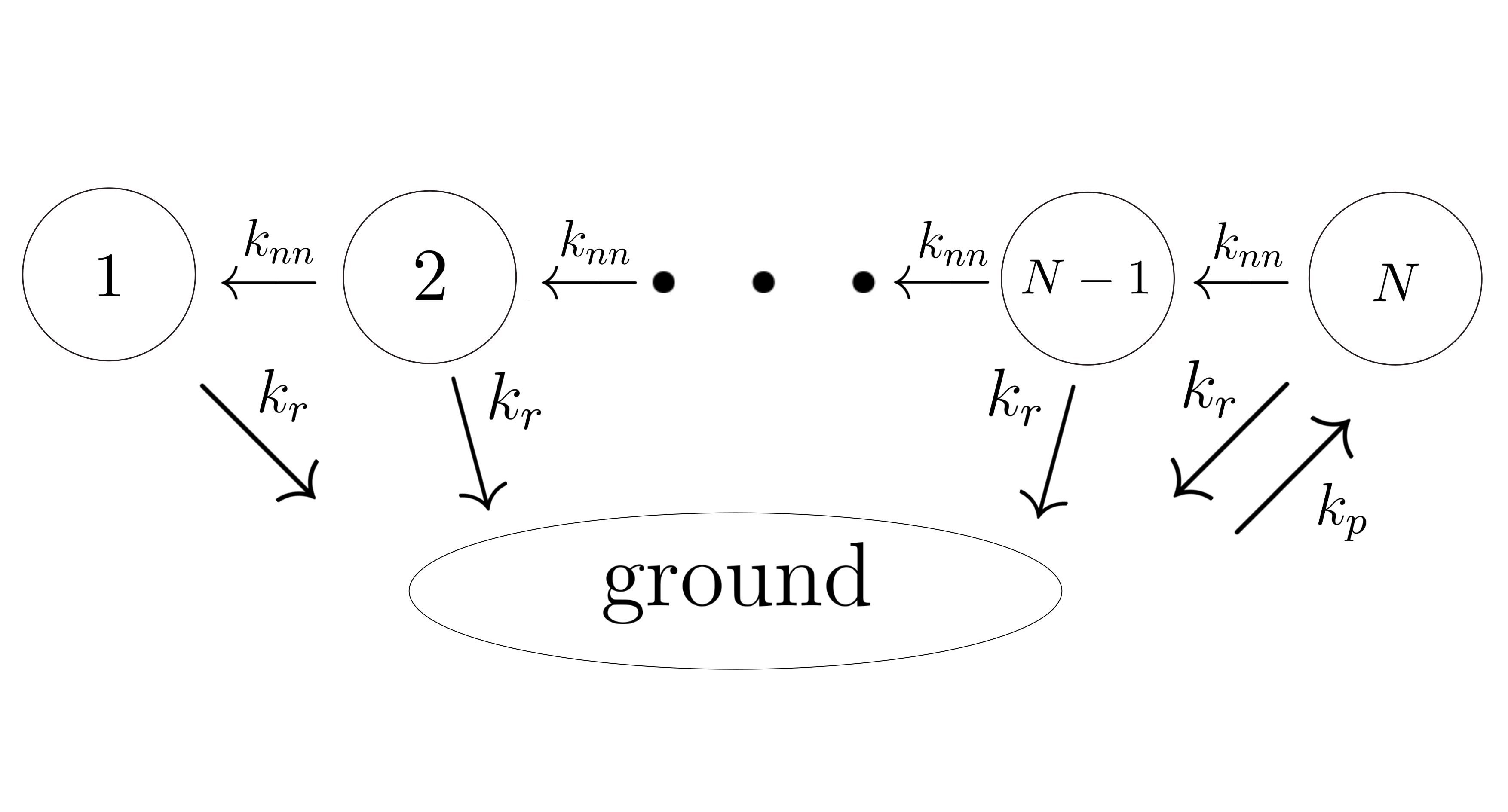}
        \caption{}
        \label{off_res_kin_model}
    \end{subfigure}%
    ~ 
    \begin{subfigure}[t]{0.5\textwidth}
        \centering
        \includegraphics[width=7.5cm]{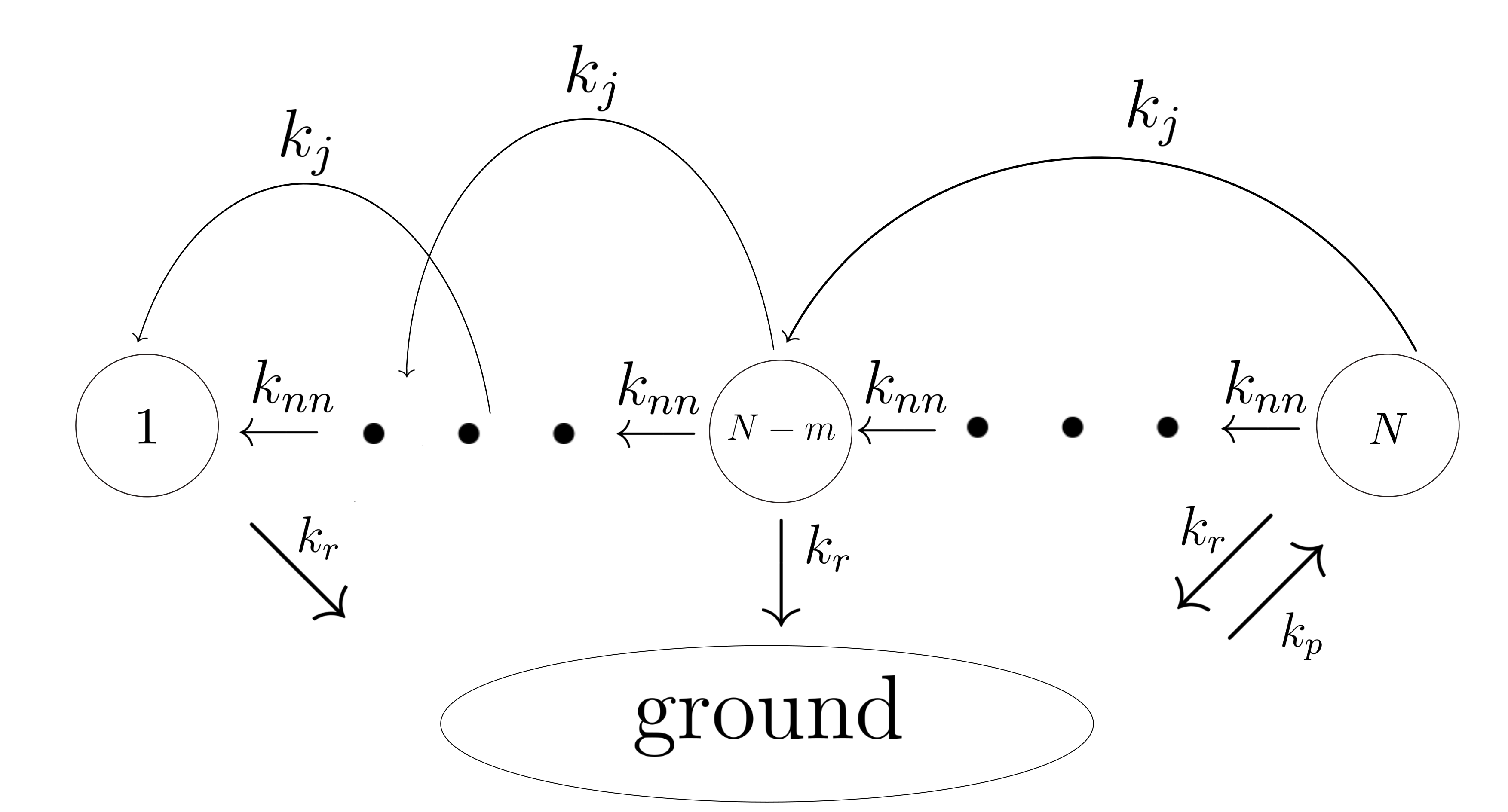}
        \caption{}
        \label{res_kin_model}
    \end{subfigure}
    \caption{Illustration of the kinetic model for (a) off-resonant $\Delta$ and (b) fractional resonant $\Delta$.}
    \label{kin_model}
\end{figure*}

To understand the implications of these qualitative considerations on exciton transport, consider a simplified kinetic model that coarse-grains over the bosonic degrees of freedom, describing the system only by the exciton populations on each site and in the ground state. A recombination rate $k_r$ and pumping rate $k_p$ are included to represent the environmental influences of the $\mathcal{L}_{r,n}$ and $\mathcal{L}_p$ operators. To incorporate the environmental parameters into the kinetic model, we set $k_r = r_r$ and $k_p = r_p$, which comes from interpreting the Lindbladian coefficients as rate constants.
\newline
\newline To incorporate the dynamics induced by the bosonic dissipation, assume that the directionality is strong enough (i.e $\Delta$ large enough) that transitions up the chain can be neglected, and that when resonant long-range mixing occurs, the strongest mixing is between the sites closest to one another (for example in Fig. \ref{on_res_delocalization}, eigenstates tend to primarily mix sites 1 and 3 and sites 3 and 5, but for most eigenstates there generally only weaker mixing between sites 1 and 5). This second assumption means that when $\Delta$ is resonant, it suffices to include only the shortest length resonant jump. These two simplifications make the steady-state solution to the kinetic model simpler to write out explicitly, but even if they are not exactly satisfied in the real Holstein model, the qualitative findings discussed below will still generally hold. More precisely, adding back the longer length jumps will shift the exciton populations somewhat further down the chain, and the reintroduction of upwards jump and nearest-neighbor transport rates will slightly shift populations upwards. The one exception is when $\Delta$ is small enough that the flow induced by $\mathcal{L}_{b,n}$ is no longer strongly biased towards the bottom sites. We will see the effect of this in a later section, but in the context in which it arises it will be straightforward to analyze by inspection.
\newline
\newline Using these two simplifying assumptions on the form of the $\mathcal{L}_{b,n}$-induced dynamics, we account for nearest-neighbor transport by allowing excitons to flow down the chain site by site with rate $k_{nn}$. Since the rate $k_{nn}$ is a consequence of the first-order mixing between adjacent sites in WS states and is induced by the bosonic dissipator $\mathcal{L}_{b,n}$, it is of the form $k_{nn} = r_b f(\Delta, \lambda, J)$ where $f$ reflects the effects of closed-system parameters on nearest-neighbor delocalization (recall $r_b$ is the inverse lifetime of the bosonic dissipation). Additionally, when $\Delta = j/m$, the model includes an $m$-site jumping process $k_{j}$. Given that this jump coupling originates from the long-range mixing and the bosonic dissipator, it is of the form $k_j = r_b g(\Delta, \lambda, J)$ where $g$ reflects the influence of closed system parameters on resonant long-range delocalization. The inclusion of the jump process also justifies the choice to coarse-grain over the bosonic degrees of freedom; the role of the bosons is still included in the dynamics by incorporating this boson-assisted jump channel where appropriate.  While we do not perform an in-depth analysis on the forms of $f$ and $g$ here, based on numerical calculations of open-system dynamics, the inferred transport rates (and hence $f$ and $g$) typically increase with the delocalization of the eigenstates (as measured using Inverse Participation Ratio \cite{benenti_principles_2018}), as expected based on how these dynamics arise from the delocalization.
\newline
\newline The kinetic model is depicted in Fig. \ref{kin_model}, where each arrow represents a linear contribution to a time derivative. We let $P_g$ denote the ground population, $P_n$ denote the population on site $n$, and $P_{ex} = \sum_{n = 1}^N P_n = 1 - P_g$ be the total excited state population. For an $N$-site chain on a $\Delta = j/m$ resonance we have:
\begin{subequations}
\begin{equation}
    \frac{dP_i}{dt} = k_{nn}[(1-\delta_{iN})P_{i+1} - (1-\delta_{i1})P_i] + \delta_{iN}k_pP_g - k_rP_i + k_{j}[\theta(N - m - i)P_{i + m} - \theta(i - m - 1)P_i]
\label{kinetic_model_equations_1}
\end{equation}
\begin{equation}
    \frac{dP_g}{dt} = k_rP_{ex} - k_{p}P_g
\end{equation}
\end{subequations}
where $\delta_{ij}$ is the Kronecker delta and $\theta(x)$ the Heaviside step function, defined by $\theta(x) = 1$ if $x \ge 0$ and $\theta(x) = 0$ otherwise. The equations for non-resonant $\Delta$ values are obtained by setting $k_{j} = 0$ in Eq. \ref{kinetic_model_equations_1}.
\newline
\newline The steady state of this model can be readily obtained in terms of normalized populations $\rho_n = P_n/P_{ex}$, as is done in Appendix \ref{appen_kin_model}. The exact form of the solution depends on the relative sizes of the chain length $N$ and jump length $m$, but all solutions show the same general behavior. For example, if  $N \ge 2m$, and the identifications $k_r = r_r$, $k_p = r_p$, $k_{nn} = r_bf(\Delta, \lambda, J)$, and $k_j = r_bg(\Delta, \lambda, J)$ are used and we introduce the ratio $R = r_b/r_r$, the steady-state $\rho_n$ are given by (c.f. Eq. \ref{long_finite_chain_pops}):
\begin{equation} \label{main_text_long_finite_chain_pops} \rho_n = \begin{cases}  \frac{1}{R(f+g) + 1} & n = N \\
    \alpha \rho_{n + 1} & N - m < n < N \\
    \gamma \rho_{n + m} + \alpha \rho_{n + 1} & m < n \le N - m \\
    \rho_n = \mu \rho_{n + m} + \nu \rho_{n + 1} & 1 < n \le m \\
    \xi \rho_{n + m} + \chi \rho_{n + 1} & n= 1 \end{cases}
\end{equation}
$\alpha, \gamma, \mu, \nu, \xi, \chi$ are dimensionless coefficients, referred to as population factors, given by:
\begin{subequations} \label{Renorm_pop_r_dep}
\begin{equation}
\rho_{N} = \frac{1}{R(f+g) + 1}
\label{renorm_pop_site_N}
\end{equation}
\begin{equation}
\alpha = \frac{f}{f + g + R^{-1}}
\end{equation}
\begin{equation}
\gamma = \frac{g}{f + g + R^{-1}}
\end{equation}
\begin{equation}
\mu = \frac{g}{f + R^{-1}}
\end{equation}
\begin{equation}
\nu = \frac{f}{f + R^{-1}}
\end{equation}
\begin{equation}
\xi = gR
\end{equation}
\begin{equation}
\chi = fR
\end{equation}
\end{subequations}
with $g = 0$ in the non-resonant case; in particular, they only depend on $f, g$, and $R$. If the jump mechanism is much faster than the nearest neighbor transport (i.e. $g \gg f$) on resonance, it is straightfoward to see from Eqns. (\ref{main_text_long_finite_chain_pops}) and (\ref{Renorm_pop_r_dep}) that sites $N - m, N - 2m, ...$ will have heightened steady-state normalized populations for resonant $\Delta$ compared to the off-resonance case. \textit{This is the resonance hypersensitivity, now predicted via kinetics}. 
\newline
\newline The kinetic model qualitatively agrees with the results in Fig. \ref{holstein_tetramer_NESS}: the spike in population at site 2 at half-integer $\Delta$ reflects the modified steady-state due to the activation of a two-site jump mechanism; similarly the spike in population on site 1 at $\Delta = j/3$ can be viewed as a consequence of the three-site jump mechanism present at those $\Delta$ values. The kinetic model indicates that as $|g|$ approaches $|f|$ the resonances should be less prominent. We expect $|g|$ to decrease exponentially with increasing denominator of $\Delta$ due to the scaling of the long-range coupling; the kinetics explains how this translates into the weakening of the NESS resonances as the denominator of $\Delta$ increases. Furthermore, one can account for the varying broadness of the resonance peaks in Fig. \ref{holstein_tetramer_NESS} by incorporating the width of the closed system resonance in the kinetic model (for example, by having the jump pathways have a smooth turn-on/turn-off as $\Delta$ passes through a resonant value); the kinetic model then shows that the closed system resonance linewidth should manifest as the NESS resonance linewidth. The decreasing width of the resonances with increasing denominator of $\Delta$ also follows from the exponential decay of the long-range coupling: the resonance behavior should occur once $\lambda(J/\Delta)^m$ is of the same order of magnitude as the smallest energy gap between WS states centered on different sites, since this is when the perturbative effects become non-negligible. This smallest gap is determined by how far $\Delta$ is from a resonant value. For smaller denominator $m$, $\lambda(J/\Delta)^m$ is more easily on the same order of magnitude, giving the broadened resonances when the denominator $m$ is small (compare the integer to half integer to $j/3$ resonances in Fig. \ref{holstein_tetramer_NESS}).
\newline
\newline The kinetic model's successful qualitative prediction suggests that the hypersensitivity of the NESS can be understood either in terms of the spectrum (location of avoided crossings) or the dynamical behavior (kinetic model) - though ultimately these two approaches are equivalent, because in the secular regime the kinetics are entirely determined by the spatial profile of the closed-system eigenstates.
\newline
\newline Before moving on to use the kinetic model as a predictive tool, consider the steps that went into making the coarse-grained kinetic model in the site basis, and when such an approach can be expected to be successful. The model only includes dynamics caused by the site populations in the product basis, and ignores the effects of off-diagonal coherences in the density operator. The ability to qualitatively represent the dynamics this way results from two factors. First, we are working in a regime of weak bath coupling where we assume that the secular approximation holds (see Appendix A), so that the energy eigenbasis population time derivative is expressed in terms of energy populations only. Second, we rely on the being able to easily characterize the spatial characteristics of the energy eigenbasis: highly localized WS states for non-resonant $\Delta$, and delocalized over sites $m$ apart when $\Delta = j/m$ is resonant. This makes it straightforward for the dynamics of the energy eigenbasis to be reexpressed in the site basis, as captured by the two exciton transport rates $k_{nn}$ and $k_j$. The applicability of this kinetic approach is therefore limited by how easily the eigenstates can be represented in the basis within which one wishes to analyze kinetics (in our case, the site basis).
\newline
\newline The other element that goes into constructing the kinetic model is the partial trace over the bosonic degrees of freedom. This was justified by inserting the resonance jump mechanism in by hand, which relies on the effects of the bosons on the eigenstates (namely, the ``re-delocalization" effect) and on the resulting effect on the bath-induced dynamics being straightforward to characterize in the site basis.
\section*{V. NESS Dependence on Bath Parameters}\label{bath_parameter_section} 
It is important to understand whether these NESS resonances will survive for different ranges of the three environmental parameters $r_r$, $r_b$ and $r_p$ in (\ref{open_sys_eom}). Depending on the physical situation, one would expect different regimes for these parameters; for example, if $\mathcal{L}_r$ represents spontaneous emission to a photon bath while $\mathcal{L}_b$ represents spontaneous emission to a phonon bath, one expects $r_b \gg r_r$ since matter-matter interaction is typically stronger than light-matter interaction. On the other hand, if light-matter coupling is enhanced (e.g. by the Purcell effect \cite{lee_gigantic_2023}) we may be able to enter regimes where $r_r \approx r_b$ or even $r_b \ll r_r$. 
\newline
\newline To understand the effect of varying the environmental parameters on the NESS, we first seek qualitative predictions from the kinetic model. From Eq. (\ref{main_text_long_finite_chain_pops}) we find recursively that the normalized populations are sums of terms of the form $(\alpha)^{a_\alpha}(\gamma)^{a_\gamma}(\mu)^{a_\mu}(\nu)^{a_\nu}(\xi)^{a_\xi}(\chi)^{a_\chi}(\rho_N)^{a_N}$, where each $a_i \in \mathbb{N} \cup \{0\}$ and 
\begin{equation} \label{bounded_greek_letters}
a_\xi = a_\chi = 0 \textrm{ if $n \neq 1$}
\end{equation} (this also holds for the other cases of $N$ and $m$ values, as shown by equations (\ref{N_eq_infty})-(\ref{N_eq_1}) in  Appendix \ref{appen_kin_model}).
Consequently, Eq. (\ref{Renorm_pop_r_dep}) shows that in the kinetic model all of the normalized populations depend on the bath parameters only through the ratio $R$ with the pumping rate $r_p$ controlling the total one-exciton population. Indeed, numerical results confirm that, as long as the secular approximation holds (Appendix \ref{appen_secular}), the NESS normalized populations
%, and more generally the dynamics of the open system Holstein model,
depend on the bath parameters only through the ratio $R = r_b/r_r$. This suggests the following physical picture: $r_p$ determines the pumping rate, only affecting the ratio of ground to one-exciton populations. $r_b$ sets a timescale for exciton motion in the one-exciton manifold since both the nearest-neighbor transport and jump rates are proportional to it, while $r_r^{-1}$ serves as a measure of exciton lifetime, so $R = r_b/r_r$ is a unitless measure of exciton lifetime, dictating the form of the NESS in the one-exciton manifold. 
\newline
\newline Since the NESS dependence on bath parameters is captured by the ratio $R$, we examine the NESS behavior in the limiting regimes of large and small $R$.
%, then consider the transition between them. 
More precisely, from Eq. (\ref{Renorm_pop_r_dep}), we see that $f^{-1}$ and $g^{-1}$ (which are unitless) define the magnitude scale of $R$. That is, $R$ being ``small"
%, ``intermediate", 
or ``large" is understood relative to the sizes of $f^{-1}$ and $g^{-1}$. It is also important to recall that the focus is on small $r_p/r_r$ in order to avoid multi-exciton effects. This restriction translates to $P_{ex}\ll1$, without restricting the value of $R$, as long as the parameters are chosen such that $r_p \ll r_r$. This means that both the large and small $R$ limits are compatible with the single-exciton restriction.
\begin{figure*}[h!]
    \centering
    \begin{subfigure}[t]{0.5\textwidth}
        \centering
        \includegraphics[width=8cm]{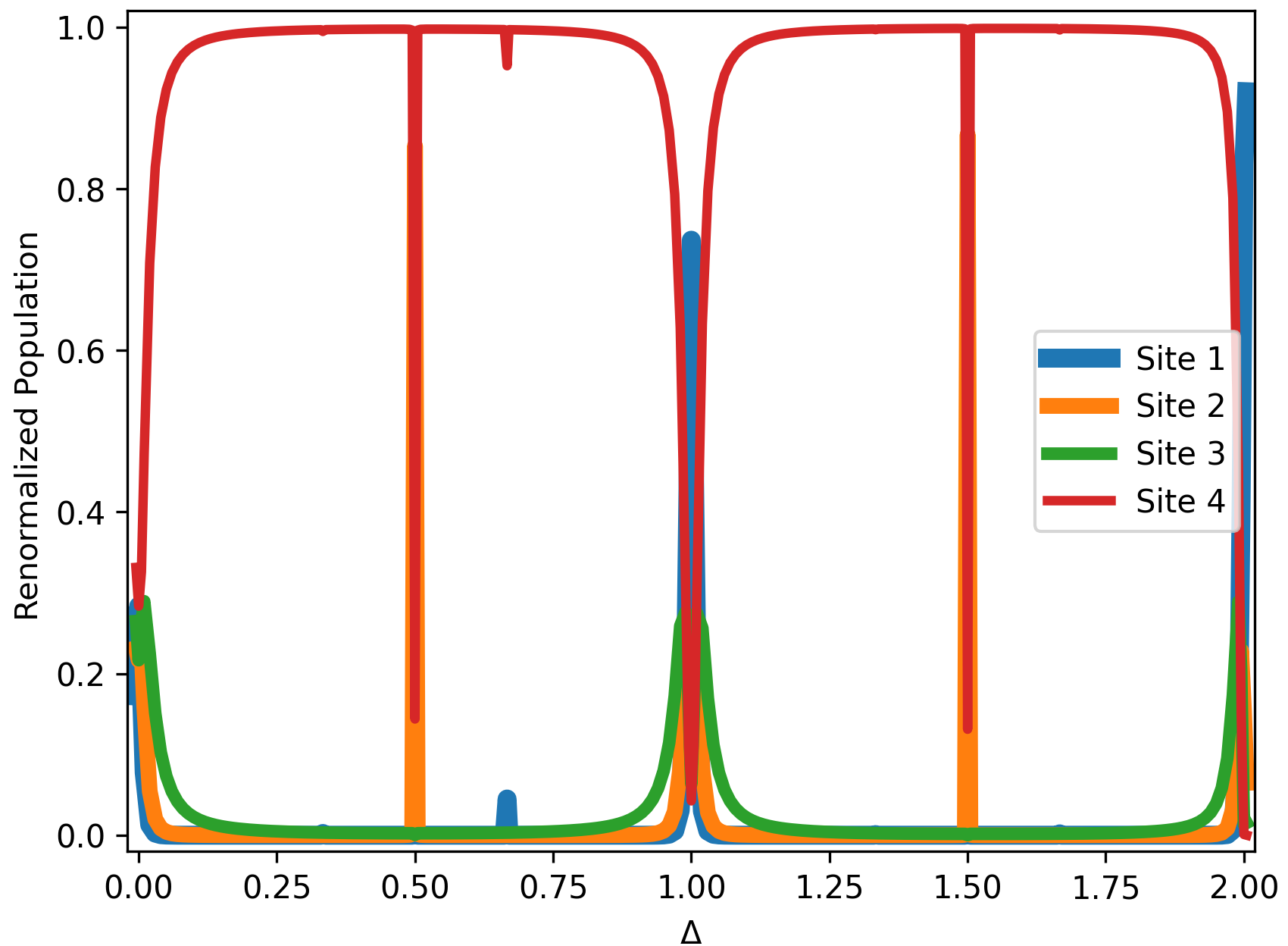}
        \caption{$R = 1$}
        \label{small_a}
    \end{subfigure}%
    ~ 
    \begin{subfigure}[t]{0.5\textwidth}
        \centering
        \includegraphics[width=8cm]{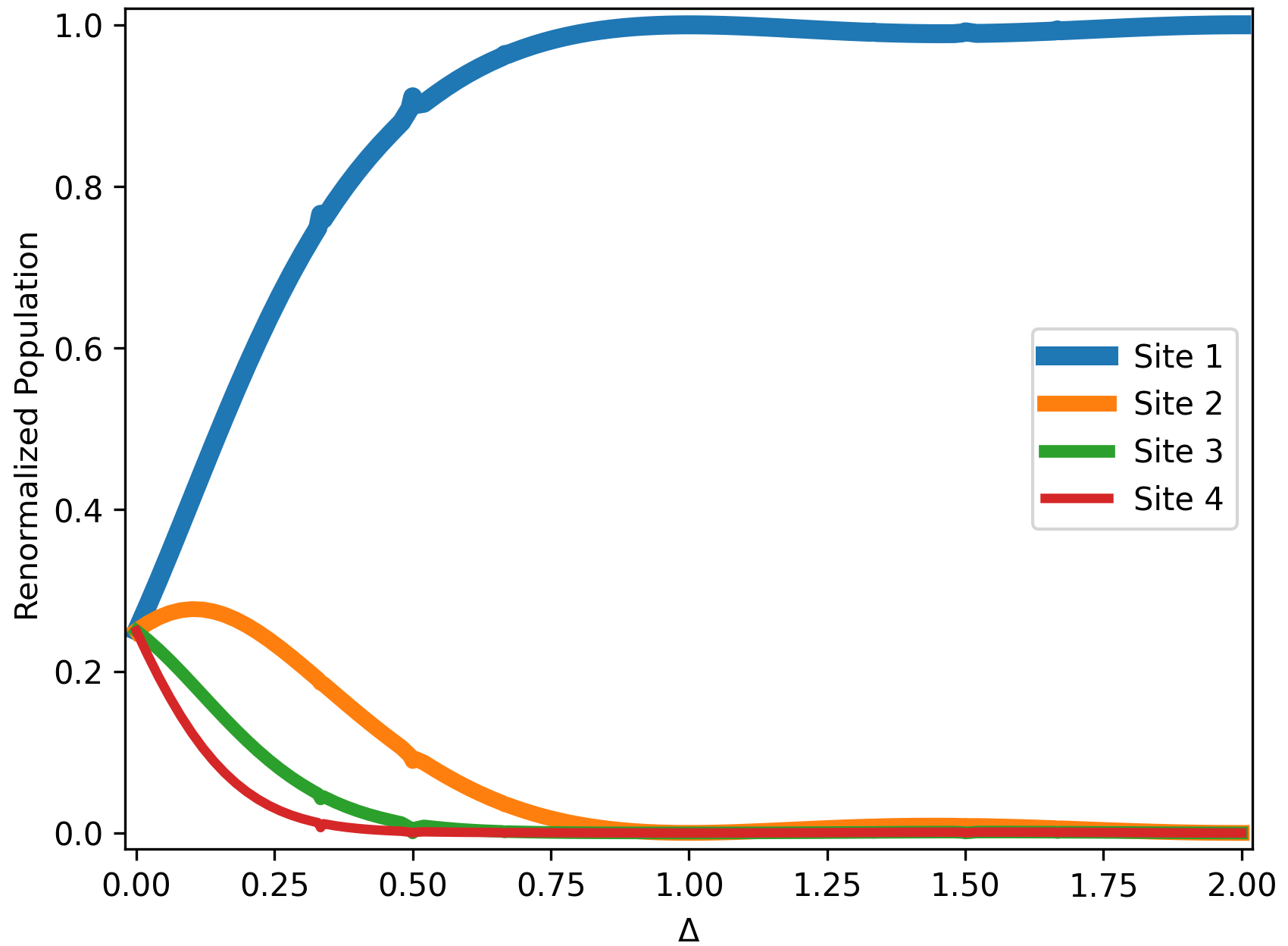}
        \caption{$R = 10^6$}
        \label{large_b}
    \end{subfigure}
    \caption{Numerically calculated normalized NESS populations for an open system Holstein tetramer; as in Fig. \ref{holstein_tetramer_NESS} $\lambda = 0.5, J = 0.01$, but for different $R$. For these values of $R$, the prominence of many of the resonances decreases. For small $R$ in (a) the population concentrates on the top site, while for large $R$ in (b) the population concentrates on the bottom site.}
    \label{R too small}
\end{figure*}
\newline
\newline In the small $R$ limit, Eq. (\ref{main_text_long_finite_chain_pops}) shows that $\rho_N \to 1$, so that almost all one-exciton population is located in the top site in both the resonant ($g \neq 0$) and non-resonant $(g = 0)$ cases. Physically, this is because the exciton lifetime is too short for even the fast resonance jump pathway to move significant population to the lower sites. This prediction is verified numerically in Fig. \ref{small_a} for the actual Holstein model (not just the simplified kinetic model), which shows the normalized NESS populations of a Holstein tetramer with the same parameters as in Fig. \ref{holstein_tetramer_NESS}, except here $R = 1$ whereas $R = 10^2$ in Fig. \ref{holstein_tetramer_NESS}. When $R = 1$ we see that for non-resonant $\Delta$ almost all population is on the top site, the weaker $\Delta = 1/3, 4/3, 5/3$ resonances have disappeared entirely, and the surviving resonances now transfer less population down the chain than when $R = 10^2$.
\newline
\newline \noindent In the large $R$ limit, i.e. long exciton lifetime, the kinetic model predicts that both resonant and non-resonant systems will have almost all of their steady-state populations concentrated on the lowest site. Indeed, from Eqns. (\ref{main_text_long_finite_chain_pops} -  \ref{bounded_greek_letters}), we see that for all sites $n$ except $n = 1$, the exciton population $\rho_n$ is a product of $\rho_N$, which tends to 0, and the population factors $\alpha, \gamma, \mu$, and $\nu$, which for fixed $f$ and $g$ are bounded above by some constant independent of $R$. Consequently, as $R$ becomes large, each $\rho_n \to 0$ for $n \neq 1$, so $\rho_1 \approx 1$ regardless of whether the resonant transport pathway is present or not. Physically, this concentration of population at the bottom of the chain indicates that the exciton lifetime is long enough that even the slower, non-resonant transport pathway can move almost all of the population to the bottom site before recombination occurs. Indeed, this is the behavior we see computationally in Fig. \ref{large_b}, which shows the NESS of the same tetramer system as in Figures \ref{holstein_tetramer_NESS} and \ref{small_a}, except now with $R = 10^6$. When $\Delta$ is large enough for highly directional transport, almost all normalized population is found on the bottom site; though the resonant transport mechanism is still present, it no longer causes a drastic change in the NESS, and the resonance is indistinguishable from the non-resonant ``background". 
\newline
\newline However, for small $\Delta$ values the kinetic model considered so far breaks down because for weak $\Delta$ the exciton flow induced by the bosonic dissipation is no longer highly directional, which was one of the simplifying kinetic model assumptions. However, it is straightforward to modify the kinetic model by adding additional nearest-neighbor and jump transport terms that go up the chain. In the limit of $\Delta \to 0$ the upward and downward rate constants become equal, so for $R$ large the NESS exciton population should be evenly distributed. This explains why the NESS tends towards all site populations being equal as $\Delta \to 0$ in Fig. \ref{large_b}.
\begin{figure}
    \centering
    \includegraphics[width=0.75\linewidth]{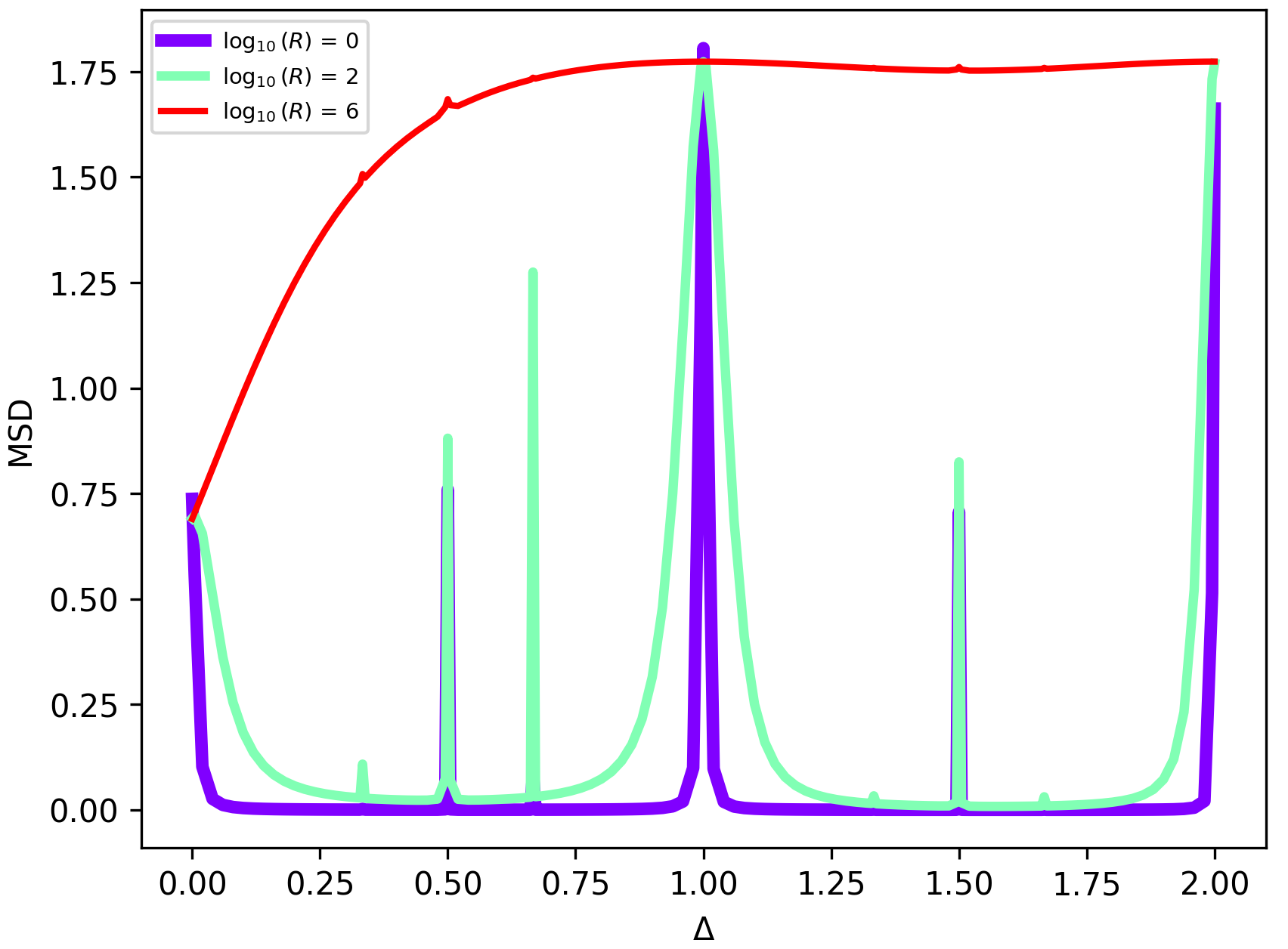}
    \caption{ Mean squared displacement for site populations on a $4$-site chain; $r_r$ and $r_p$ were maintained at $10^{-9}$ and $10^{-10}$ respectively, while $r_b$ was varied to produce the different $R$ values, shown in the upper corner.}
    \label{MSD_vs_R}
\end{figure}
\newline
\newline \noindent To see how this $R$ dependence manifests in the actual populations, we numerically calculated the mean squared displacement (MSD) of the steady state site populations from the top site \cite{kessing_long-range_2022},  MSD $= \sum_{n = 1}^N (N - n)^2 P_n = P_{ex}\sum_{n = 1}^N (N - n)^2 \rho_n$, as a function of $\Delta$ for various values of $R$ obtained by holding $r_r$ and $r_p$ fixed and varying $r_b$. Interest in the MSD is motivated by the physical interpretation of the Holstein model as representing the transport of excitons that are initially localized on the top site; the MSD then acts as a measure of the transport properties of the system in the steady-state limit. The numerically calculated steady state MSD is shown in Fig. \ref{MSD_vs_R} for the same tetramer system as in Figures \ref{holstein_tetramer_NESS} and \ref{R too small}. The dependence of the fractional resonances on $R$ is reflected in the MSD, with the effect being most prominent in an intermediate $R$ regime, as was the case with the normalized populations. This is expected from the kinetic model since $P_{ex} = k_p/(k_r + k_p) = r_p/(r_r + r_p)$, so for fixed $r_r$ and $r_p$ the actual and normalized populations should just be multiples of one another and therefore show the same resonance behavior. Indeed, for the tetramer example, we find that $P_{ex}$ is approximately constant for all $\Delta$ and $R$ plotted, showing the expected independence of $\Delta$ and $r_b$. The value of $P_{ex}$ can be changed by varying $r_r$ or $r_p$, which causes a corresponding change in the magnitude of the MSD, with the lineshape remaining the same.
\section*{VI. Conclusion}
We have shown the resonance hypersensitivity in the non-equilibrium steady state (NESS) of a linearly biased Holstein chain subject to environmental influences in the weak intersite coupling ($J \ll \Delta$) regime. This hypersensitivity manifests at fractional values of the linear bias $\Delta$, where the NESS exciton populations are drastically altered such that the top site and the sites located $m,2m,...$ ($m$ being the denominator of the fractional $\Delta$) from the top are populated, with intermediate sites left relatively unpopulated. This hypersensitivity correlates with the locations of the avoided crossings in the closed system spectrum or alternatively can be predicted via kinetic arguments. The kinetic approach models both system and environmental factors, consequently providing a natural framework to understand how varying environmental parameters affect NESS hypersensitivity. In particular, kinetic arguments intuitively explain the apparent disappearance of the resonance hypersensitivity in the limits of small and large $R = r_b/r_r$, a dimensionless order parameter that quantifies the lifetime of the charge/energy carriers. Although the insight here is from the Holstein model, this connection between the NESS and the dynamics of the system is expected to hold in general, suggesting that one can predict the existence and parameter dependence of NESS hypersensitivity based on relatively straightforward kinetic arguments. 
\newline
\newline This work also shows that a large degree of control can be exhibited over the Holstein model NESS via a combination of tuning the linear bias $\Delta$ and the ``normalized lifetime" $R$. Generally, closed-system resonances lead to hypersensitive NESS transport, i.e. a high on-off ratio, when the open-system parameters are such that resonance-enabled channels contribute significantly to the overall transport efficacy, which is the case for the intermediate exciton lifetime regime. Two notable counter-examples are the $R\rightarrow0$ and $R\rightarrow\infty$ limits. In the former case the lifetime is too short for either channel to contribute significantly to NESS transport, while in the latter the lifetime is longer even than the time it takes to reach the harvest site using non-resonant channels.
\newline
\newline These findings could be useful in engineering the performance of open quantum systems operating in the steady-state regime; for example selective transport to a specific position could be achieved by a synergistic tuning of closed- and open-system parameters ($\Delta$ and $R$ in this study). However, for real systems we note that such dichotomy of system parameters is generally unavailable and their cross-correlations become less tractable, for example when there are degrees of freedom that are in the intermediate coupling regime with time scales comparable to transport in question. A general theoretical framework that accounts for this is left for future work.
\subsection*{Acknowledgements}
This material is based upon US Air Force Office of Scientific Research (AFOSR) under grant FA9550-20-1-0354 and support from the Natural Sciences and Engineering Research Council of Canada (NSERC).
\section*{Appendix}
\renewcommand{\theequation}{A-\arabic{equation}}
\renewcommand{\theHequation}{A\arabic{equation}}
\setcounter{equation}{0}
\renewcommand{\thefigure}{A-\arabic{figure}}
\renewcommand{\theHfigure}{A\arabic{figure}}
\setcounter{figure}{0}
\renewcommand\thesubsection{\Alph{subsection}}
\setcounter{subsection}{0}
\subsection{Validity of Secular Approximation} \label{appen_secular}
 Recent studies \cite{dodin_population_2024} have demonstrated that the bath generates coherences in systems with $\ge 4$ bath-induced transitions between energy eigenstates; however, no conclusion could be made on the size of the coherences, since it differs for each system. Bath-induced coherences, if large enough, can lead to dynamics that differ significantly from the secular regime, rendering the secular approximation inaccurate. Alternatively, there are regimes for which the secular approximation is preferred over the nonsecular \cite{dodin_secular_2018}. This appendix shows that it is safe to apply the secular approximation to the Holstein model considered in this work: even if the coherences impact the NESS, the form of the bath interaction ensures that the qualitative results seen in the main text remain unchanged.
 \newline
 \newline Starting from the full evolution of the density operator (where the sum over $\alpha$ denotes the sum over all Lindblad operators in Eq. (\ref{open_sys_eom})):
\begin{equation}
\label{open_sys_eom_appen}
\dot{\rho} = -i[H, \rho] + \sum_{\alpha} \left(L_\alpha\rho L_\alpha^\dagger - \frac{1}{2}\{L_\alpha^\dagger L_\alpha, \rho\}\right)
\end{equation}
and working in the interaction picture with $U(t) = \exp(-iHt)$ and $A^{(I)}(t) = U^\dagger(t) A(t) U(t)$, one finds that the entries of $\rho^{(I)}$ in the closed-system energy eigenbasis evolve according to (using $L_{\alpha, ij}$ for the $ij$ entry of $L_\alpha$):
\begin{equation} \label{full_eom_energy_eigenbasis}
\begin{split} 
    \frac{d\rho^{(I)}_{ij}}{dt} &= \sum_{\alpha} \left(\sum_{k, m} L_{\alpha, ik}^{(I)}(t)(L^\dagger_{\alpha, mj})^{(I)}(t)\rho^{(I)}_{km}(t) - 
    \frac{1}{2}\sum_{k} \left[ (L^\dagger L)^{(I)}_{\alpha, ik}(t)\rho^{(I)}_{kj}(t) - (L^\dagger L)^{(I)}_{\alpha, kj}(t)\rho^{(I)}_{ik}(t) \right] \right) \\ &= \sum_{\alpha, k,m} e^{-i(\omega_{ik} + \omega_{mj})t}L_{\alpha,ik}L^\dagger_{\alpha, mj}\rho^{(I)}_{km}(t) - \frac{1}{2}\sum_{\alpha, k} \left(e^{-i\omega_{ik}t}(L^\dagger L)_{\alpha, ik}\rho^{(I)}_{kj}(t) - e^{-i\omega_{kj}t}(L^\dagger L)_{\alpha, kj}\rho^{(I)}_{ ik}(t)\right)
\end{split}
\end{equation}
In particular, using $\rho^{(I)}_{ii}(t) = \rho_{ii}(t)$, one finds that the populations in the energy eigenbasis evolve according to:
\begin{equation} \label{pop_evol_non_sec}
\begin{split}
    \dot{\rho_{ii}}(t) = \sum_{\alpha, k,m} e^{-i(\omega_{mk})t}L_{\alpha, ik}L^*_{\alpha, im}\rho^{(I)}_{km}(t) - \sum_{\alpha, k} \Re[e^{-i\omega_{ik}t}(L^\dagger L)_{\alpha, ik}\rho^{(I)}_{ki}(t)]
\end{split}
\end{equation}
To understand when the evolution of populations and coherences decouple, we first estimate the size of $|\dot{\rho}^{(I)}_{ij}(t)|$. Set $M = \sup_{\alpha, i,j}\{|L_{\alpha, ij}|\}$ and note that each $|\rho^{I}_{ij}(t)|$ is $\mathcal{O}(1)$ because $|\rho^{I}_{ij}(t)| = |\rho_{ij}(t)| \le \sqrt{\rho_{ii}(t)\rho_{jj}(t)} \le 1$ (using $|\rho_{ij}(t)| \le \rho_{ii}(t)\rho_{jj}(t)$ for a density operator $\rho$ \cite{cohen-tannoudji_quantum_2019}). Because the eigenstates in the closed system are approximately spatially localized WS states with well-defined total bosonic excitation number (off resonance) or delocalized only over a few nearby sites (on resonance), each state is only coupled to a few others through the Lindbladians $L_{\alpha}$, because the Lindbladians considered in this work (e.g. $L_{b,n} = \sqrt{r_b}a_n$) can only couple eigenstates that overlap spatially. Consequently, the summation over $\alpha, k$, and $m$ in (\ref{full_eom_energy_eigenbasis}) should only introduce approximately $\mathcal{O}(1)$ non-zero terms into the summation.
%(or at the very least, the number of terms it introduces does not diverge in the limit of infinite system size and therefore just introduces a constant prefactor to the magnitude of $|\rho^{I}_{ij}(t)|$). 
Altogether, one sees from Eq. (\ref{full_eom_energy_eigenbasis}) that $|\dot{\rho}^{(I)}_{ij}(t)|$ is $\mathcal{O}(M^2)$.  
\newline
\newline Therefore,  if $M^2/\omega \ll 1$, then for any term $e^{i\omega t}\rho^{I}_{km}(t)$ in (\ref{pop_evol_non_sec}), one has $e^{i \omega t}\rho^{I}_{km}(t) \approx e^{i \omega t}\rho^{I}_{km}(0)$ over a period of oscillation of $e^{i \omega t}$, hence the integral of $e^{i\omega t}\rho^{I}_{km}(t)$ approximately vanishes and its contribution to the time evolution (\ref{pop_evol_non_sec}) can be neglected. For the model used in this work, $M^2$ is on the order of $r = \max\{r_p, r_b, r_r\}$. %[CC: what about the wavefunction part of L? Perhaps one can argue that if the corresponding wavefunction amplitudes are small then there wouldn't have been much coherence to start with? I think in our case most states are completely localized so worries there but things might get funny within the minibands, again.]
Therefore, if $\Omega = \min_{i \neq j} \{\omega_{ij}\}$ is the smallest energy gap in the closed system, then in the limit where $r \ll \Omega$, one can drop all oscillatory terms from (\ref{pop_evol_non_sec}), yielding the secular equations of motion for the eigenstate populations:
\begin{equation} \label{secular_pop_eom_appen}
\begin{split}
    \dot{\rho_{ii}}(t) = \sum_{\alpha, k} |L_{\alpha, ik}|^2\rho_{kk}(t) - \sum_{\alpha, k} |L_{\alpha, ki}|^2\rho_{ii}(t) \; \; \; \textrm{(Full Secular)}
\end{split}
\end{equation}
In this limit the closed-system evolution only affects the population dynamics by determining the identity of the energy eigenstates; the phase factor oscillations that are characteristic of closed-system evolution in the energy eigenbasis are too fast to affect the steady-state populations. 
\newline
\newline Dropping the oscillatory terms from (\ref{full_eom_energy_eigenbasis}), the secular evolution of the energy eigenbasis coherences becomes:
\begin{equation}
    \dot{\rho_{ij}}^{(I)}(t) = \sum_{\alpha} \left(L_{\alpha, ii}L_{\alpha, jj}^* - \frac{1}{2}\left((L^\dagger L)_{\alpha, ii} + (L^\dagger L)_{\alpha, jj}\right)\right)\rho_{ij}^{(I)}(t) \; \; \; \textrm{(Full Secular)}
\end{equation}
Using the Cauchy-Schwarz and AM-GM inequalities, as well as the particular form of the Lindbladians considered in this work, it is straightforward to check that Re$(\dot{\rho_{ij}}^{(I)}(t)/\rho_{ij}^{I}(t)) < 0$, hence in the secular regime, the coherences should decay exponentially and become negligible in the NESS. Thus, in the secular limit $r \ll \Omega$, the NESS can be calculated by solving for the steady state populations using Eq. (\ref{secular_pop_eom_appen}), and then setting the coherences to 0, as is done in the main text. 
\newline
\newline However, given that that Holstein spectrum at fractional resonance is a collection of minibands with each band containing nearly degenerate states, the secular picture requirement that $r \ll \Omega$ is quite restrictive, and it is important to understand the effects of reintroducing non-secular terms. Consider, therefore, the partial secular regime, where $r$ is larger than the frequencies between states within a miniband, but smaller than the frequencies between states in different bands, which is on the order of the bosonic quanta.
%,making this partial secular regime realistic
Retaining the non-secular terms in (\ref{full_eom_energy_eigenbasis}) between states within the same miniband, the population evolution becomes (absorbing the complex exponential factors into the coherences to write them in the Schrodinger picture):
\begin{equation} \label{partial_secular_eom}
\begin{split}
\dot{\rho_{ii}}(t)&= \underbrace{\sum_\alpha \sum_{k \not\in \textrm{band}} |L_{\alpha, ik}|^2\rho_{kk} - |L_{\alpha, ki}|^2\rho_{ii}}_{\textcircled{1}} \\ &+ \underbrace{\sum_\alpha \sum_{k \in \textrm{band}, m \in \textrm{band}} L_{\alpha, ik}L^*_{\alpha, im}\rho_{km}(t)}_{\textcircled{2}} - \underbrace{\sum_\alpha \sum_{k \in \textrm{band}, k \neq i} \Re[(L^\dagger L)_{\alpha, ik}\rho_{ki}(t)]}_{\textcircled{3}}
\end{split}
\end{equation}
\newline where $k \in \textrm{band}$ means states $k$ in the same miniband as state $i$. In the partial secular regime, coherences can redistribute populations within a miniband (terms \textcircled{2} and \textcircled{3}), while the transitions between bands is purely secular (term \textcircled{1}). To examine the intraband coherent transport, first note that the $L_{b,n}$, $L_{r,n}$, and $L_p$ only have matrix elements between different minibands, so term \textcircled{2} vanishes. Additionally, note that $(L^\dagger_{p}L_p)_{ik} = r_p\delta_{i,g}\delta_{k,g}$ where $g$ denotes the ground state, so the $L_p$ contribution to term \textcircled{3} also vanishes.  Finally, $L^\dagger_{b,n}L_{b,n} = r_b N_n$, where $N_n = a_n^\dagger a_n$ is the bosonic number operator for mode $n$, and $L_{r,n}^\dagger L_{r,n} = r_r M_n$ where $M_n = c_n^\dagger c_n$ is the exciton occupation number.%, i.e. it can only couple states in the same band that have spatial overlap. 
\subsubsection*{Off-Resonant $\Delta$}
For off-resonant $\Delta$ in the limit of small $J/\Delta$, the closed-system eigenstates are Wannier-Stark (WS) states multiplied by some number state, hence the energy eigenbasis also approximately diagonalize $L^\dagger_{b,n}L_{b,n}$ and $L_{r,n}^\dagger L_{r,n}$ since they are multiples of the boson and exciton number operators, respectively. Therefore, $L_{b,n}^\dagger L_{b,n}$ approximately does not couple states within the miniband for off-resonant $\Delta$, so their contribution to term \textcircled{3} vanishes and the off-resonance partial secular population evolution reduces to a secular form:
\begin{equation}
\begin{split}
    \dot{\rho_{ii}}(t) = \sum_{\alpha, k} |L_{\alpha, ik}|^2\rho_{kk}(t) - \sum_{\alpha, k} |L_{\alpha, ki}|^2\rho_{ii}(t) \; \; \; \textrm{(Partial Secular; Off-Resonance)}
\end{split}
\end{equation}
In particular, this means that the partial secular off-resonance NESS eigenstate populations are the same as the full secular case. 
\newline
\newline However, even if the off-resonance NESS energy eigenbasis coherences are unmodified by the reintroduction of intraband coherences, non-zero NESS coherences could still in principle affect the NESS site populations and Mean Square Displacement (MSD) $= \sum_{n = 1}^N (N - n)^2 P_n$, which were the main focus in the main text. Indeed, taking the off-resonance energy eigenstates to be products $|k\rangle_{WS}\otimes|\vec{m} \rangle = | k, \vec{m} \rangle_{WS}$ (a valid approximation off-resonance since the large gaps between bands prevents the perturbation $V$ from inducing strong mixing), the population $P_j$ on site $j$ becomes (letting $\vec{m}, \vec{n}, \vec{n}'$ index the bosonic configurations):
\begin{equation}
\begin{split}
\label{off_res_sum}
    P_j &= \sum_{\vec{m}} \langle j, \vec{m}|\rho| j, \vec{m}\rangle \\
    &= \sum_{k_{WS}, k_{WS}' \vec{m},\vec{n}, \vec{n}'} \langle k_{WS}, \vec{n}|\rho|k_{WS}' \vec{n}'\rangle \langle j,\vec{m}|k_{WS}, \vec{n}\rangle\langle k_{WS}', \vec{n}'|j, \vec{m} \rangle \\
    &= \sum_{k_{WS}, k_{WS}', \vec{m}} J_{j -k}(2J/\Delta)J_{j - k'}(2J/\Delta)\langle k_{WS}, \vec{m}|\rho|k_{WS}', \vec{m}\rangle 
\end{split}
\end{equation}
where the last line used the orthogonality of the bosonic modes and the definition (\ref{ws_def}) of the WS states. In the above sum, the $k_{WS} = k_{WS}'$ terms are the contributions to $P_j$ from the eigenstate populations $\langle k_{WS}, \vec{m}|\rho|k_{WS}, \vec{m}\rangle $, which were shown to be unaffected by the reintroductiong of intraband coherences. Therefore, any modifications to the site populations and the MSD comes from the $\langle k_{WS}, \vec{m}|\rho|k_{WS}', \vec{m}\rangle$ for $k_{WS} \neq k_{WS}'$, i.e. the ``intersite" coherences between states localized on different sites. Assuming uniqueness of the NESS, it is straightfoward to show that within the partial secular regime these intersite coherences are negligible in the NESS, and thus the site populations and MSD --- the key observables studied in the main text --- are unaffected by intraband coherences for off-resonant $\Delta$. Indeed, returning to expression (\ref{full_eom_energy_eigenbasis}) to derive the partial secular equations for coherence evolution, the second term (summation over $\alpha$ and $k$ alone) reduces to a purely secular form by similar arguments to those used for the population evolution. However, because the first term (the sum over $\alpha,k$, and $m$) includes the differences between energy gaps in the exponential, it could in principle include non-secular contributions if there are two pairs of state $(|i\rangle, |k\rangle)$ and $(|j\rangle, |m\rangle)$, such that $\omega_{ik} - \omega_{jm}$ is small relative to $r$. We are interested in the case where $|i\rangle$ and $|j\rangle$ are localized on different WS states. Then the evolution takes the form:
\begin{equation} 
\begin{split} 
\frac{d\rho^{(I)}_{ij}}{dt} &= \sum_{\alpha} \left(L_{\alpha, ii}L_{\alpha, jj}^* - \frac{1}{2}\left((L^\dagger L)_{\alpha, ii} + (L^\dagger L)_{\alpha, jj}\right)\right)\rho_{ij}^{(I)}(t) \\ &+ \sum_{\alpha, \{(k,m) \mid \omega_{ik} - \omega_{jm} < r , (k,m) \neq (i,j)\}} e^{-i(\omega_{ik} - \omega_{jm})t}L_{\alpha,ik}L^\dagger_{\alpha, mj}\rho^{(I)}_{km}(t)
\end{split}
\end{equation}
Because the $L_{b,n}$ are the only Lindbladians that couple one-exciton states, they are the only contribution to the second term. But these $L_{b,n}$ do not affect the spatial profile of the state, so for $L_{b,n, ik}$ and $L^\dagger_{b,n, jm}$ to be non-zero, $|i\rangle$ must be localized on the same WS state as $|k \rangle$, and $|j\rangle$ must be localized on the same WS state as $|m \rangle$. %(note we are neglecting the effects of the perturbation (\ref{WS_state_coupling}) on the spatial profile of the states). 
Altogether, this means that the evolution of coherences between states with spatial profiles WS states $|k_{WS}\rangle$ and $|k_{WS}'\rangle$ only couples to other coherences between those same WS states (but different bosonic excitations). In particular, since there is nothing generating these spatially delocalized coherences between different WS states, if there are no initial intersite coherences, there can not be any at later times. This implies that if the NESS is unique (and all initial conditions tend to the steady state as $t \to \infty$), it must also have negligible intersite coherences and thus the site populations and MSD are unaffected by Eqn. (\ref{off_res_sum}). Altogether, we see that for off-resonant $\Delta$, the eigenstate population evolution is identical to the secular limit and the non-zero NESS coherences do not affect the MSD and site populations, so the secular arguments used to study the NESS hypersensitivity in the main text should hold approximately in the non-secular regime when $\Delta$ is off-resonant, so long as $r$ remains much smaller than the interband energy gap.
\newline
\newline This expectation that the secular approximation accurately captures the off-resonant NESS was tested by numerically calculating steady states for $\Delta = 0.99, \lambda = 0.2, J = 0.01$ with and without the secular approximation. For these parameters only one steady state for the Liouvillian could be found numerically, and therefore the above argument that the coherences do not affect the NESS populations applies. For these parameters, the smallest interband energy gap was $\mathcal{O}(10^{-2})$ while typical intraband energy gaps were between $10^{-7}$ and $10^{-5}$ (setting the minimum frequency $\Omega$); it was found that the secular and non-secular NESS site populations were within $1\%$ of each other for $r \le \mathcal{O}(10^{-4})$; since this magnitude of $r$ is larger than the intraband energy gap this suggests that as expected from the above discussion, secular dynamics for off-resonant $\Delta$ are accurate even beyond the strict limit of $r \ll \Omega$. It is worth noting that even once the numerical results of the secular and non-secular NESS site populations began to differ, general trends such as the MSD's dependence on $R$ were the same for both secular and non-secular NESS.
\subsubsection*{Resonant $\Delta$}
For resonant $\Delta$, there is strong mixing between different unperturbed eigenstates, and we therefore expect $M_n$ and $N_n$ (defined below Eq. (\ref{partial_secular_eom})) to induce non-negligible coupling within a band. Therefore, the partial secular population evolution takes the form:
\begin{equation} 
\begin{split}
\dot{\rho}_{ii}
(t) &= \begin{cases} \!\begin{aligned}%[b]
       & \sum_\alpha \left( \sum_{k \not\in \textrm{band}} |L_{\alpha,ik}|^2\rho_{kk}(t) - |L_{\alpha,ki}|^2\rho_{ii}(t)\right) 
       \\ & - \sum_{n}  \sum_{k \in \textrm{band}, k \neq i} \textrm{Re}\left(\left(r_b(N_n)_{ik} + r_r(M_n)_{ik}\right)\rho_{ki}(t)\right) 
    \end{aligned} & \textrm{(one-exciton states)}
     \\ \begin{aligned} \sum_\alpha \left( \sum_{k} |L_{\alpha,ik}|^2\rho_{kk}(t) - |L_{\alpha,ki}|^2\rho_{ii}(t)\right) \end{aligned} & \textrm{(ground state)}
\end{cases}
\\ \; \; \; 
 &\textrm{(Partial Secular; On-Resonance)}
\end{split}
\end{equation}
A similar evaluation of the intraband coherence evolution shows that they can, in principle, be non-zero at long times, so the coherence terms in the above expression do not vanish in the NESS. Therefore the intraband coherences induce extra population transfer within each band, redistributing the band's population amongst its states. The presence of non-zero coherences additionally gives rise to interference effects that can further alter the NESS site populations. While this will change the populations and MSD of the NESS, we do not expect the qualitative results of the secular regime to change. This is because the NESS site populations and MSD depend primarily on the bath-induced dynamics \textit{between} bands, not the distribution \textit{within} a band (which is ultimately the reason that the the kinetic picture that coarse-grains over the bands is conceptually useful). Even in the partial secular regime, the transport between bands is still achieved through secular dissipation by $L_b$ that moves population into lower-energy minibands, with resonant transport occurring at fractional $\Delta$ due to the energy-matching condition and associated delocalization within each band. The competition of timescales (and the consequent $R$-dependence of MSD) should also still be present: the NESS distribution of population among the different minibands should still be determined by the relative rates of the bosonic dissipation that induces population flow down the minibands and the exciton recominbation that siphons population into the ground state.
\newline
\newline This prediction that the resonance NESS is qualitatively captured by the secular approximation was tested by numerically comparing secular and non-secular NESS site populations for $\Delta = 1.00, \lambda = 0.2, J = 0.01$, for which the smallest interband gap is $\mathcal{O}(1)$, and typical intraband gap is between $\mathcal{O}(10^{-7})$ to $\mathcal{O}(10^{-5})$. The secular and non-secular populations agree (within $1\%$) for $r < \mathcal{O}(10^{-6})$; the smaller threshold compared to the off-resonant case described suggests that the non-secular transport within a band plays a more significant role on resonance, as expected from the analytic arguments. However, even for $r$ greater than the threshold value, the general features of the secular regime, namely the occurrence of fractional resonances and the lineshape of the $R$ dependence of the NESS populations, were still present, suggesting that the findings in the work do extend beyond the secular regime. 

\subsection{Analytical Steady-State Solution of Kinetic Models} \label{appen_kin_model}
Here we solve for the steady state of the on-resonance kinetic model (Eq. (\ref{kinetic_model_equations_1})) depicted in Fig. \ref{res_kin_model} for a chain of arbitrary length $N$; setting $k_{j} = 0$ in Eq. (\ref{kinetic_model_equations_1}) yields the off-resonance steady state, depicted in Fig. \ref{off_res_kin_model}. Let $P_g$ denote the ground state population, $P_{n}$ the population on site $n$, and $P_{ex} = P_1 + ... + P_{N}$ be the total one-exciton population. We have:
\begin{equation}
    \frac{d P_g}{dt} = k_r(P_{ex}) - k_p(P_g),
\end{equation}
so in the steady state:
\begin{equation} \label{ground_ss}
    P_g = \frac{k_r}{k_p}P_{ex}.
\end{equation}
Since the populations sum to unity, $1 - P_{ex} = P_g$, which combined with Eq. (\ref{ground_ss}) gives:
\begin{equation} \label{one_ex_ss}
P_{ex} = \frac{k_p}{k_r + k_p}
\end{equation}
and:
\begin{equation} \label{a_4}
P_g = \frac{k_r}{k_r + k_p}
\end{equation}
Since $\frac{dP_N}{dt} = -(k_{nn} + k_{j} + k_r)P_N + k_{p}P_g$, in the steady state we have:
\begin{equation} \label{top_ss}
P_N = \frac{k_p}{k_{nn} + k_{j} + k_r}P_g
\end{equation}
We introduce the normalized one-exciton populations $\rho_n \coloneqq P_n/P_{ex}$ ($1 \le n \le N)$, so that the $\rho_n$ represent the relative populations of each site in the one-exciton manifold, with $\sum \rho_n = 1$. Equations (\ref{one_ex_ss}), (\ref{a_4}), and (\ref{top_ss}) give:
\begin{equation} \label{top_renorm}
\rho_N = \frac{k_r}{k_{nn} + k_{j} + k_r}
\end{equation}
Considering the normalized populations hence allows us to remove the steady-state dependence on $k_p$; all of the dependence on $k_p$ is contained in equations (\ref{ground_ss}) and (\ref{one_ex_ss}). The remaining $\rho_i$ can be obtained recursively; due to the jump term $k_{j}$ the exact results depend on the chain length.  The result is:
\begin{itemize}
\item $N = \infty$ (with sites labeled from $-\infty$ to some top site $N_{\textrm{top}}$):
\begin{equation}
    \rho_n = \begin{cases} \frac{k_r}{k_{nn} + k_{j} + k_r} & n = N_{\textrm{top}} \\
    \alpha \rho_{n + 1} & N_\textrm{top} - m < n < N_\textrm{top} \\
    \gamma \rho_{n + m} + \alpha \rho_{n + 1} & n \le N_\textrm{top} - m
    \end{cases}
    \label{N_eq_infty}
\end{equation}
\item $N > 2m$:
\begin{equation} \label{long_finite_chain_pops}
    \rho_n = \begin{cases} \frac{k_r}{k_{nn} + k_{j} + k_r} & n = N \\
    \alpha \rho_{n + 1} & N - m < n < N \\
    \gamma \rho_{n + m} + \alpha \rho_{n + 1} & m < n \le N - m \\
    \rho_n = \mu \rho_{n + m} + \nu \rho_{n + 1} & 1 < n \le m \\
    \xi \rho_{n + m} + \chi \rho_{n + 1} & n = 1
\end{cases}
\end{equation}
\item $2m \ge N > m$:
\begin{equation}
    \rho_n = \begin{cases}
        \frac{k_r}{k_{nn} + k_{j} + k_r} & n = N \\
        \alpha \rho_{n + 1} & m < n < N \\
        \rho_n = \nu \rho_{n + 1} & N - m < n \le m \\
        \rho_n = \mu \rho_{n + m} + \nu \rho_{n + 1} & 1 < n \le N - m \\
    \xi \rho_{n + m} + \chi \rho_{n + 1} & n = 1
    \end{cases}
\end{equation}
\item $m \ge N > 1$:
\begin{equation}
    \rho_n = \begin{cases} \frac{k_r}{k_{nn} + k_{j} + k_r} & n = N \\
    \rho_n = \nu \rho_{n + 1} & 1 < n \le N - m \\
    \rho_n = \chi \rho_{n + 1} & n = 1
    \end{cases}
\end{equation}
\item $N = 1$:
\begin{equation}
    \rho_n = 1
    \label{N_eq_1}
\end{equation}
\end{itemize}
with $\alpha, \gamma, \mu, \nu, \xi$ and  $\chi$ defined by:
\begin{subequations} \label{greek_letters}
\begin{equation}
\alpha = \frac{k_{nn}}{k_{j} + k_{nn} + k_r}
\end{equation}
\begin{equation}
\gamma = \frac{k_{j}}{k_{j} + k_{nn} + k_r}
\end{equation}
\begin{equation}
    \nu = \frac{k_{nn}}{k_{nn} + k_r}
\end{equation}
\begin{equation}
    \mu = \frac{k_{j}}{k_{nn} + k_r}
\end{equation}
\begin{equation}
\chi = \frac{k_{nn}}{k_r}
\end{equation}
\begin{equation}
\xi = \frac{k_{j}}{k_r}
\end{equation}
\end{subequations}
In particular, upon performing the substitutions $k_p = r_p, k_r = r_r, k_{nn} = r_b f(\Delta, \lambda, J), k_j = r_b g(\Delta, \lambda, J)$, we recover Eq. (\ref{Renorm_pop_r_dep}) from the main text.
\newline
\newline
 The non-resonant case ($k_{j} = 0$) is particularly simple, with:
\begin{subequations} \label{non_res_analytical_soln}
\begin{equation}
\rho_n = \nu^{N - n}(\frac{k_r}{k_{nn} + k_{j} + k_r}) = \nu^{N -n}\rho_{N} \textrm{ (for $n > 1$)}
\end{equation}
\begin{equation}
\rho_1 = \chi\nu^{N  - 2}\rho_{N}
\end{equation}
\end{subequations}
\bibliographystyle{unsrt}
\bibliography{references}
\end{document}